\documentstyle[epsfig]{mn}
\ifCUPmtlplainloaded \else
 \def\umu{\mu}
 \def\upi{\pi}
 
\fi
\title[The 1996 Outburst of GRO~J1655--40: Spectral Results] {The 1996
Outburst of GRO~J1655--40: The Challenge of Interpreting the
Multi-wavelength Spectra}
\author[R.I.~Hynes, et al.]
       {R.I.~Hynes$^1$, C.A.~Haswell$^{1,2}$, 
        C.R.~Shrader$^3$, W.~Chen$^3$, Keith~Horne$^4$, 
        \newauthor E.T.~Harlaftis$^4$, K.~O'Brien$^4$, C.~Hellier$^5$ 
        and R.P.~Fender$^{1,6,7}$, \\
        $^1$Astronomy Centre, University of Sussex, Falmer, Brighton BN1 9QH\\
        $^2$Columbia Astrophysics Laboratory, Columbia University, 
            538 West 120th Street, New York, NY 10027, USA\\
        $^3$Goddard Space Flight Center, Greenbelt, MD 20771, USA\\
        $^4$School of Physics and Astronomy, University of St. Andrews, 
            North Haugh, St. Andrews, Fife KY16 9SS\\
        $^5$Department of Physics, Keele University, Keele ST5 5BG\\
        $^6$Astronomical Institute `Anton Pannekoek', University of
       	    Amsterdam, The Netherlands\\
	$^7$Center for High Energy Astrophysics, Kruislaan 403, 
            1098 SJ, Amsterdam, The Netherlands}
\begin{document}
%
%
\newcommand{\novasco}{GRO\,J1655--40}
\newcommand{\novaper}{GRO\,J0422+32}
\newcommand{\novamus}{X-ray Nova Muscae 1991}
\newcommand{\sigmasco}{$\sigma$~Sco}
%
%
\newcommand{\HST}  {\textit{HST}}
\newcommand{\XTE}  {\textit{RXTE}}
\newcommand{\RXTE} {\textit{RXTE}}
\newcommand{\GRO}  {\textit{CGRO}}
\newcommand{\CGRO} {\textit{CGRO}}
\newcommand{\ASCA} {\textit{ASCA}}
\newcommand{\ROSAT}{\textit{ROSAT}}
%
%
\newcommand{\AAT}{AAT}
\newcommand{\CTIO}{CTIO}
%
%
\newcommand{\HI}   {H\,\textsc{i}}
\newcommand{\HII}  {H\,\textsc{ii}}
\newcommand{\HeI}  {He\,\textsc{i}}
\newcommand{\HeII} {He\,\textsc{ii}}
\newcommand{\HeIII}{He\,\textsc{iii}}
\newcommand{\CI}   {C\,\textsc{i}}
\newcommand{\CII}  {C\,\textsc{ii}}
\newcommand{\CIII} {C\,\textsc{iii}}
\newcommand{\CIV}  {C\,\textsc{iv}}
\newcommand{\KI}   {K\,\textsc{i}}
\newcommand{\OIII} {O\,\textsc{iii}}
\newcommand{\NIII} {N\,\textsc{iii}}
\newcommand{\FeII} {Fe\,\textsc{ii}}
\newcommand{\SiIII}{Si\,\textsc{iii}}
\newcommand{\SiIV} {Si\,\textsc{iv}}
%
%
\newcommand{\EBV}{E(B-V)}
\newcommand{\Rv} {R_{\rm V}}
\newcommand{\Av} {A_{\rm V}}
%
%
\newcommand{\lam}   {$\lambda$}
\newcommand{\lamlam}{$\lambda\lambda$}
\maketitle
%
%
\newcommand{\comm}[1]{\textit{[#1]}}
%
%
\begin{abstract}

  We report on the results of a multi-wavelength campaign to observe
  the soft X-ray transient (SXT) and superluminal jet source \novasco\
  in outburst using \HST, \RXTE\ and \CGRO\ together with ground based
  facilities.  This outburst was qualitatively quite different to
  other SXT outbursts and to previous outbursts of this source.  The
  onset of hard X-ray activity occurred very slowly, over several
  months and was delayed relative to the soft X-ray rise.  During this
  period, the optical fluxes {\em declined} steadily.  This apparent
  anti-correlation is not consistent with the standard disc
  instability model of SXT outbursts, nor is it expected if the
  optical output is dominated by reprocessed X-rays, as in persistent
  low mass X-ray binaries.
  
  Based on the strength of the 2175\,\AA\ interstellar absorption
  feature we constrain the reddening to be $\EBV=1.2\pm0.1$, a result
  which is consistent with the known properties of the source and with
  the strength of interstellar absorption lines.  Using this result we
  find that our dereddened spectra are dominated by a component
  peaking in the optical with the expected $\nu^{1/3}$ disc spectrum
  seen only in the UV.  We consider possible interpretations of this
  spectrum in terms of thermal emission from the outer accretion disc
  and/or secondary star, both with and without X-ray irradiation, and
  also as non-thermal optical synchrotron emission from a compact
  self-absorbed central source.  In addition to the prominent \HeII\ 
  4686\,\AA\ line, we see Bowen fluorescence lines of \NIII\ and
  \OIII, and possible P~Cygni profiles in the UV resonance lines,
  which can be interpreted in terms of an accretion disc wind.  The
  X-ray spectra broadly resemble the high-soft state commonly seen in
  black hole candidates, but evolve through two substates.
  
  Taken as a whole, the outburst dataset cannot readily be interpreted
  by any standard model for SXT outbursts.  We suggest that many of
  the characteristics could be interpreted in the context of a model
  combining X-ray irradiation with the limit cycle disc instability,
  but with the added ingredient of a very large disc in this long
  period system.
\end{abstract}
%
%
\begin{keywords}
accretion, accretion discs -- binaries: close -- stars: individual: Nova Sco
1994 (GRO J1655--40) -- ultraviolet: stars -- X-rays: stars
\end{keywords}
%
%
\section{Introduction}
Soft X-ray transients (SXTs), also referred to as X-ray novae,
\cite{TS96} are a class of low-mass X-ray binaries (LMXBs) in which
long periods of quiescence, typically decades, are punctuated by very
dramatic X-ray and optical outbursts, often accompanied by radio
activity as well.  The most promising models for explaining the
outbursts invoke the thermal-viscous limit cycle instability
previously developed for cataclysmic variables \cite{C93}.  These have
met with some success in explaining the properties of the outbursts
\cite{CCL95} but there remain difficulties (e.g.\ Lasota, Narayan \&
Yi 1996).  Compared to cataclysmic variables, an important effect that
must be included in models of SXTs is X-ray irradiation of the disc
and/or the secondary star.  Irradiation of the disc will change its
temperature structure \cite{TMW90} and may induce delayed reflares
(Chen, Livio \& Gehrels 1993, Mineshige 1994).

The SXT \novasco\ was discovered in 1994 July when \GRO\ Burst and
Transient Source Experiment (BATSE) observed it in outburst at a level
of 1.1\,Crab in the 20--200\,keV energy band \cite{Ha95}.  Since then
it has undergone repeated outbursts to a similar level and shown
itself to be a very atypical SXT.  The outburst history from 1994--5
has been summarised by Tavani et al.\ \shortcite{T96}, who draw
attention to the contrast between the 1994 outbursts which were
radio-loud with apparent superluminal jets observed (Tingay et al.\
1995, Hjellming \& Rupen 1995) and the 1995 outbursts at a similar
X-ray flux as in 1994, but radio-quiet.  The optical flux from
\novasco\ is not as well documented, but Orosz, Schaefer \& Barnes
\shortcite{OSB95} note that optical brightening does not always
accompany X-ray outbursts.

After a period of apparent quiescence from late 1995 to early 1996,
\novasco\ went into outburst again in late 1996 April \cite{R96}.
Orosz et al.\ \shortcite{O97} observed an optical rise leading the
X-ray rise detected by \XTE\ by about 6~days. They suggested that this
initial behaviour was consistent with the limit-cycle instability.
The subsequent X-ray behaviour, however, was not as expected.  The
soft X-ray flux (2--10\,keV), as followed by the \XTE\ All Sky Monitor
(ASM) remained at an approximately constant level for more than
4~months, though with considerable short term variability while the
hard X-ray flux (20--200\,keV) as monitored by \GRO\ BATSE was
observed to rise very slowly, not reaching its peak until 4~months
after the initial dramatic increase in the soft flux.

During this period we carried out a series of simultaneous \HST\ and
\XTE\ visits, backed up by ground based observations and \GRO\ BATSE
data.  We present here our spectral analysis.  A subsequent paper will
address timing issues.  First we summarise the current state of
knowledge on the properties of \novasco: the context in which we
interpret our results.
\subsection{System parameters}
\label{ParameterSection}
The system parameters of \novasco\ are the best known of any SXT.
They are summarised in Table~\ref{ParameterTable}.  Hjellming \& Rupen
\shortcite{HR95} estimate the distance from a kinematic model of the
jets to be $3.2\pm0.2$\,kpc.  We also have a lower limit from
observations of the 1420\,MHz interstellar absorption \cite{T95} of
3.0\,kpc and an upper limit of 3.5\,kpc obtained by the method of
Mirabel and Rodr\'{\i}guez \shortcite{MR94}.  The latter assumes that
we can correctly identify the proper motions of the two jets relative
to the central source and then only involves the requirement that
these proper motions are produced by material moving at no more than
the speed of light.  These two constraints support the distance
estimate of Hjellming \& Rupen.  Other parameters are taken from Orosz
\& Bailyn \shortcite{OB97} who model the quiescent light curve at a
time when the disc is estimated to contribute less than 10 per cent of
the V band light.  Their deduced mass of $7.02\pm0.24$\,M$_{\sun}$
makes it clear that the compact object in this system is a black hole.
We note that an independent parameter determination by van der Hooft
et al.\ \shortcite{vdH97} yields values consistent with those of Orosz
\& Bailyn \shortcite{OB97}, although with larger uncertainties.
\begin{table}
\caption{Adopted parameters for \novasco.}
\label{ParameterTable}
\begin{center}
\begin{tabular}{lc}
\noalign{\smallskip}
\hline
\noalign{\smallskip}
Distance              & $3.2\pm0.2$\,kpc                    \\
Period                & $2.62157\pm0.00015$\,d              \\
Mass function         & $3.24\pm0.09$                       \\
Inclination           & $69\fdg50\pm0\fdg08$      \\
Mass ratio            & $2.99\pm0.08$                       \\
Primary mass          & $7.02\pm0.22$\,M$_{\sun}$          \\
\noalign{\smallskip}  
\hline
\end{tabular}
\end{center}
\end{table}
%
%
\section{Observations and data reduction}
\subsection{\HST}
\label{HSTSection}
Our spectra were obtained by \HST\ on five separate visits from 1996
May~14 to July~22 using the Faint Object Spectrograph (FOS) in RAPID
mode \cite{K95}.  Table \ref{ExpTable} details the exposures obtained.
We use the ephemeris of Orosz \& Bailyn \shortcite{OB97} to calculate
spectroscopic phases at the midpoint of each exposure.

To examine the spectra, we combined the multiple images produced by
the RAPID mode into a single image.  In addition, where several images
were taken with the same disperser, we took an average weighted by
exposure time, rebinning higher substep data where necessary.

There are several regions of our spectra for which source count rates
become comparable to those in unexposed regions; we must pay careful
attention to these background levels to avoid systematic errors in our
calibration.  There are actually two issues here.  Firstly the FOS
detectors are always subject to a particle induced dark count.  This
is modelled and subtracted by the FOS pipeline data processing
software but is generally underestimated by up to $\sim30$ per cent
\cite{K95}; the actual amount is a function of geomagnetic position
and varies over the course of an exposure.  This becomes a problem at
the short wavelength end of the PRISM spectra.  We correct these cases
by measuring the count rate in unexposed regions and rescaling the
pipeline's background to this value on an exposure by exposure basis.
When the G160L grating is used to observe a very red source such as
\novasco\ there can also be a significant amount of light scattered
from higher orders.  The pipeline measures the residual count rate in
unexposed regions of the detector after subtracting the modelled
background and subtracts this residual count from the whole spectrum,
i.e.\ it assumes no variation in count rate with pixel number.

In reality, the excess above the {\em model} background is likely to
be due to a combination of underestimating the background and extra
scattered light.  We therefore compared the results of the pipeline
reduction (which assumes all the excess is due to scattered light with
a flat spectrum) to rescaling the model background to the unexposed
count rate.  Since the model background is nearly flat, the
differences are small and negligible compared to the uncertainty in
the background level.  As the excess is much larger for G160L than for
other dispersers, for the final analysis we assume that in this case
it is dominated by scattered light and accordingly use the pipeline
reduction.

\begin{figure}
\begin{center}
\epsfig{angle=90,width=3in,height=2in,file=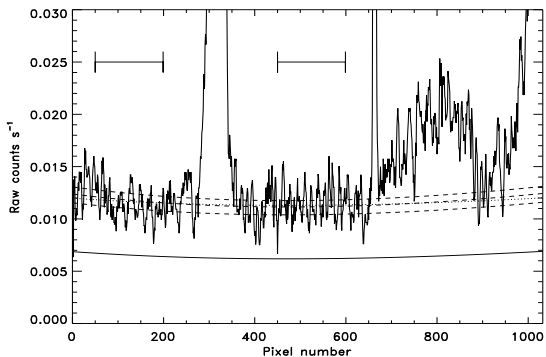}
\caption{Raw count rates in our first G160L spectrum, over the whole
  detector, smoothed with a 5 pixel boxcar.  We see the zeroth order
  spectrum centred at pixel 320.  The first order spectrum runs from
  pixels 638--1032.  Most prominently we see the geocoronal Ly$\alpha$
  emission around pixel 660 and the 2175\,\AA\ interstellar absorption
  feature around pixel 925.  We also show with the smooth solid line
  the model background calculated based upon the geomagnetic position,
  and with the dashed lines the same scaled up to match the unexposed
  regions (including our adopted lower and upper limits on this
  scaling.)  The dotted line shows the model background with a fixed
  (wavelength independent) constant added; it does not differ
  significantly from the rescaled background.  The two adopted
  unexposed regions are indicated.  They span pixels 50--199 and
  450--599 respectively.}
\label{G160CountsFig}
\end{center}
\end{figure}

We estimate the uncertainty in the background level by examining how
the count rate varies between 50~pixel bins over the unexposed region;
we are concerned here with real deviations between different parts of
the detector, not simply the statistical uncertainty in setting the
level of this model.  Figure~\ref{G160CountsFig} shows one of our raw
G160L spectra, together with the pipeline background and our rescaled
background.

For the other gratings used there are no unexposed regions in which
the background can be checked.  The count rate in G400H is much
greater than the background for all pixels so this is not an issue.
The short wavelength end of the G270H data ($\la 2600$\,\AA) probably
is affected and so should be used with caution.  The G190H data is
very sensitive to the assumed background for all pixels and disagreed
strongly with the G160L data. Since this was only a short exposure, we
chose to reject this data.

Spectra obtained using the PRISM have a non-uniform dispersion, which
becomes very low at the red end of each spectrum (232\,\AA\ per pixel
at the far end of the lowest resolution red PRISM spectra).  Under
these circumstances, a small error in centring the star can lead to a
large error in the wavelength and hence flux calibration.  In the case
of blue PRISM data, comparison with the red PRISM allows us to
estimate the point at which this becomes significant; this varies from
visit to visit from 2900--4500\,\AA.  For the red PRISM, we do not
have any internal checks of the calibration.  We compared our spectra
with near simultaneous AAT observations (see Sect.~\ref{AATObs}).  The
flux levels are consistent to within the intrinsic variability between
different nights of the AAT spectra and the shape generally agree well
up to $\sim7000$\,\AA.  We do see a sharp drop off at the longest
wavelengths which we attribute to miscentring, so we adopt cut-offs on
a case by case basis ranging from 7300--7650\,\AA.

For the first visit there were some mismatches at the overlaps of
spectra, which could be due to calibration uncertainties (estimated at
5-10 per cent) or intrinsic variability in the source.  The blue and
red PRISM and G400H data all agreed well, so we took these as a
reference and then multiplied the G270H data by 0.92 and the G160L
data by 1.06 to ensure consistency in the 3235--3276\,\AA\ and
2437--2509\,\AA\ regions respectively. As noted above, the relative
calibration of G160L and G270H is very uncertain due to the background
contribution to the G270H data.

As can be seen from Fig.~\ref{G160CountsFig}, the G160L spectra show a
strong geocoronal Ly$\alpha$ emission feature, with negligible source
counts shortwards of this.  We therefore ignore data below 1255\,\AA.
We also observe a very weak second order Ly\,$\alpha$ at 2435\,\AA\
and exclude pixels contaminated by this.  Finally, we note that the
red PRISM data from the first visit (only) showed pronounced a dip
from 3735.8--3991.4\,\AA\ that was not matched by features in the
grating spectra.  This region is about 20 pixels wide (the width that
would be affected by a single dead diode) so we also exclude these
pixels.

\begin{figure*}
\epsfig{angle=90,width=6in,height=4in,file=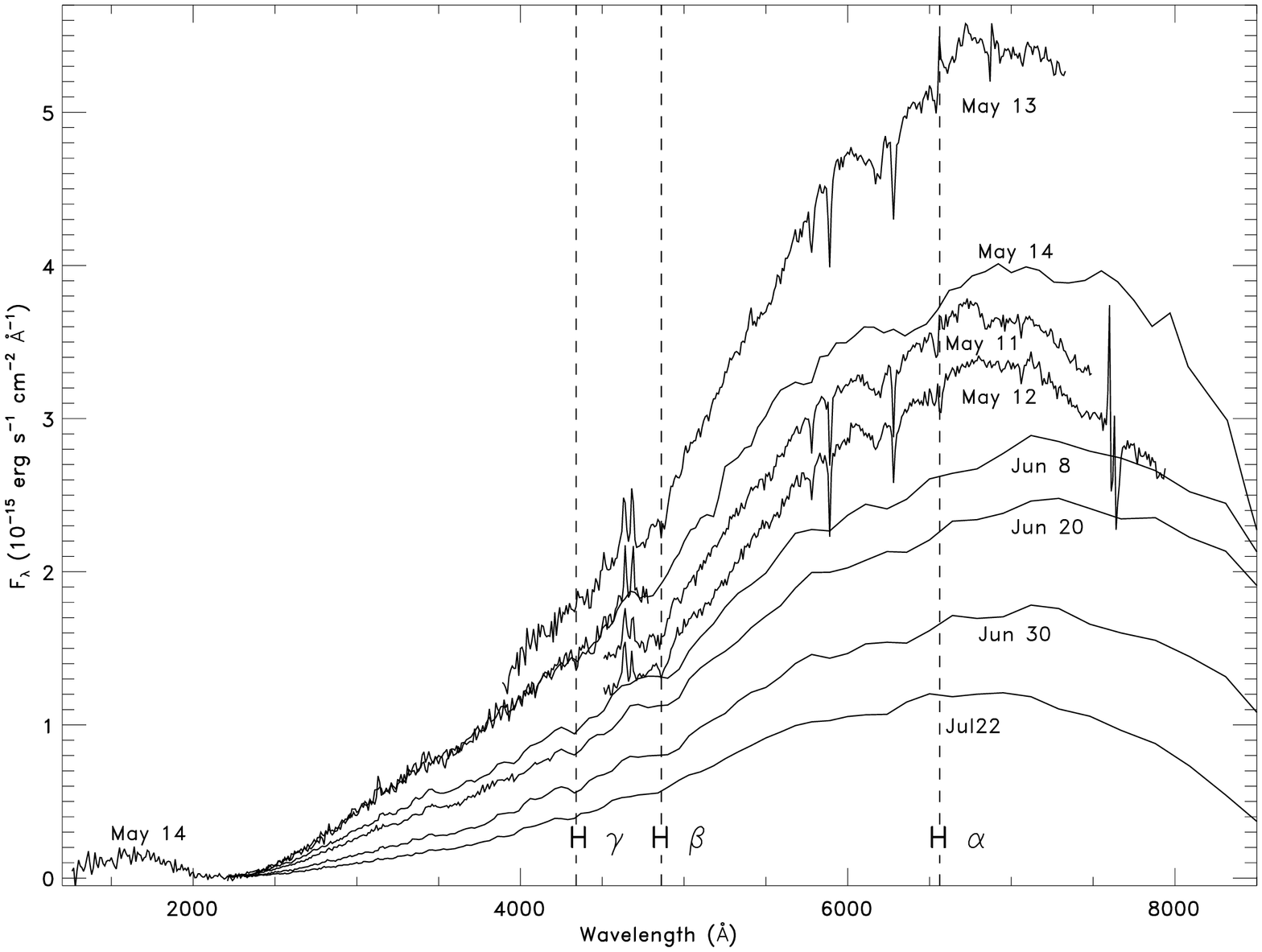}
\caption{Composite spectra for all \HST\ and \AAT\ observations.  Spectra have 
  been rebinned to 10\,\AA\ for clarity.  The long wavelength data has
  been retained for comparison here, but was truncated for subsequent
  analysis.  See text for details.}
\label{CompSpecFig}
\end{figure*}

We produced a composite PRISM spectrum for each visit together with a
separate grating spectrum (for the first visit only) which agrees very
well with the PRISM spectrum.  The resultant composite spectra are
shown in Fig. \ref{CompSpecFig} together with the \AAT\ spectra
described in Sect.~\ref{AATObs}.
\begin{table*}
\begin{minipage}{150mm}
\caption{Log of 1996 {\HST} observations of \novasco.  The PRISM was
  used with both the blue (BL) and red (RD) detector.  For the
  gratings, the resolution, $\lambda / \Delta\lambda$ is about 250 for
  G160L and 1300 for G190H, G270H, G400H.  Spectroscopic phase zero is
  when the secondary star has maximum positive radial velocity.}
\label{ExpTable}
\begin{center}
\begin{tabular}{llcccccc}
\hline
\noalign{\smallskip}
Date
& Disperser  & Start & End  & Integrating & Midpoint      & Wavelength  & 
                                                               Reciprocal \\
&            & time  & time & time        & spectroscopic & range (\AA) & 
                                                               Dispersion \\
&            & (UT)  & (UT) & (s)         & phase         &  & 
                                                      (\AA\,pixel$^{-1}$) \\
\noalign{\smallskip}
\hline
\noalign{\smallskip}
May 14 
& G190H      & 00:07:23 & 00:14:35 &  365.4   & 0.35 & 1573--2330  & 0.4   \\
& G160L      & 00:19:50 & 00:46:58 & 1574.9   & 0.36 & 1138--2507  & 3.5   \\
& PRISM (BL) & 01:42:24 & 01:47:55 &  280.0   & 0.38 & 1547--5940  & --   \\
& G160L      & 01:53:40 & 02:23:32 & 1733.0   & 0.38 & 1138--2507  & 3.5   \\
& G160L      & 03:18:27 & 03:36:02 &  893.2   & 0.41 & 1138--2509  & 1.7   \\
& G160L      & 03:39:37 & 03:47:11 &  438.5   & 0.41 & 1138--2507  & 3.5   \\
& PRISM (BL) & 03:52:51 & 03:58:55 &  308.5   & 0.41 & 1547--5940  & --   \\
& G270H      & 04:56:21 & 05:10:00 &  692.9   & 0.43 & 2221--3277  & 0.5   \\
& G400H      & 06:32:11 & 06:39:40 &  379.6   & 0.46 & 3235--4781  & 0.7   \\
& PRISM (RD) & 08:08:30 & 08:14:40 &  313.2   & 0.48 & 1621--8849  & --   \\
\noalign{\smallskip}
May 20 & \multicolumn{5}{l}{\it Target acquisition failed} \\
\noalign{\smallskip}
May 27 & \multicolumn{5}{l}{\it Target acquisition failed} \\
\noalign{\smallskip}
June 8
& PRISM (RD) & 12:48:53 & 13:06:28 &  893.2   & 0.09 & 1620--8887  & --    \\
& PRISM (RD) & 14:21:57 & 15:03:01 & 2383.2   & 0.12 & 1620--8771  & --    \\
& PRISM (BL) & 16:01:31 & 16:41:42 & 2332.2   & 0.15 & 1501--5903  & --    \\
& PRISM (BL) & 17:41:22 & 18:17:33 & 2099.9   & 0.17 & 1501--5903  & --    \\
\noalign{\smallskip}
June 20
& PRISM (RD) & 11:36:02 & 11:46:41 &  617.4   & 0.65 & 1620--8771  & --    \\
& PRISM (RD) & 12:34:59 & 12:51:35 &  963.4   & 0.67 & 1620--8771  & --    \\
& PRISM (RD) & 12:55:16 & 13:12:51 &  994.9   & 0.67 & 1620--8771  & --    \\
& PRISM (BL) & 14:12:25 & 14:51:56 & 2293.8   & 0.70 & 1500--5903  & --    \\
& PRISM (BL) & 15:52:53 & 16:34:28 & 2413.1   & 0.72 & 1500--5903  & --    \\
\noalign{\smallskip}
June 30
& PRISM (RD) & 13:01:46 & 13:12:25 &  617.4   & 0.49 & 1620--8771  & --    \\
& PRISM (RD) & 14:03:31 & 14:20:07 &  963.4   & 0.50 & 1620--8771  & --    \\
& PRISM (RD) & 14:23:48 & 14:41:23 &  994.9   & 0.51 & 1620--8771  & --    \\
\noalign{\smallskip}
July 22
& PRISM (RD) & 07:10:05 & 07:27:40 &  893.2   & 0.79 & 1620--8887  & --    \\
& PRISM (RD) & 08:43:07 & 09:26:23 & 2511.5   & 0.81 & 1620--8771  & --    \\
\hline
\end{tabular}
\end{center}
\end{minipage}
\end{table*}
\subsection{Ground based spectroscopy}
\label{AATObs}
We observed \novasco\ on the nights of 1996 May~11--13 with the
Anglo-Australian Telescope (\AAT). These observations formed part of a
co-ordinated campaign to observe the variability of \novasco\ with
high time resolution optical spectroscopy and X-ray timing with \RXTE\
(to appear in a subsequent paper) The RGO spectrograph was used in
conjunction with the 270R grating and Tektronix CCD. This gave a mean
reciprocal dispersion of 3.4\,\AA\,pixel$^{-1}$.  Exposure times and
wavelength ranges varied over the three nights, due to changes in
instrumental set-up. See Table \ref{AATTable}.

The spectra were extracted from the raw images with the optimal
extraction algorithm of Horne \shortcite{H86}, using suitable fits for
the sky lines. The calibration used subroutines within the
\textsc{molly} spectroscopic data reduction package. Spectra were
wavelength calibrated using 3rd order polynomial fits to CuAr arc
calibration spectra. The flux calibration used wavelength dependent
fits to tabulated data for the flux star LTT~4364 \cite{H94},
including telluric line extinction effects. The final stage of the
calibration used wide-slit exposures of the flux star to make an
approximate correction for light loss in the object frames due to the
narrower slit width used.

We do observe a near linear drop off in the spectra at the longest
wavelengths that does not agree well with the \HST\ spectra.  We
interpret this as due to differential refraction moving the red part
of the image off the slit and truncate the spectra at 7100\,\AA\
accordingly.
\begin{table*}
\begin{minipage}{150mm}
\caption{Log of \AAT\ spectroscopic observations.}
\label{AATTable}
\begin{center}
\begin{tabular}{lcccc}
\noalign{\smallskip}
\hline
\noalign{\smallskip}
Date   & Start time & Duration & Spectroscopic & Useful wavelength \\
       & (UT)       & (s)      & phase         & range (\AA)       \\
\noalign{\smallskip}
\hline
\noalign{\smallskip}
May 11 & 13:34:24   & 300      & 0.42          & 4500--7495        \\
May 12 & 13:39:44   & 200      & 0.81          & 4498--7947        \\
May 13 & 13:31:47   & 500      & 0.19          & 3884--7341        \\
\noalign{\smallskip}  
\hline
\end{tabular}
\end{center}
\end{minipage}
\end{table*}
\subsection{\XTE\ data}
We observed \novasco\ with the \RXTE\ Proportional Counter Array (PCA)
at 6 separate epochs, 4 of which coincide with our \HST\ pointings.
Table~\ref{XTETable} shows the observing log in which we list the
start/end time in UT and the duration of the on-source exposure.  The
total exposure time was 40.74\,ks.  We obtained data at various time
resolutions, as one goal of this program was to search for
cross-correlations between rapid variability in optical/UV/X-ray
bands.

The PCA, which consists of 5 gas proportional counter modules,
provides spectral coverage over the 2--60\,keV range. For a complete
discussion of the \RXTE\ instrumentation see Jahoda et al.\ 
\shortcite{J96}.  We used two standard PCA data modes together with a
16 energy channel mode binned at 4-ms time resolution and two single
channel, single bit modes at 62-$\umu$s.  We constructed spectra from
the `standard mode' data with 128 energy channels over the bandpass.
Typical count rates were $\sim10^{4}$\,s$^{-1}$, so statistical errors
were very small. Response matrices and estimated background count
spectra were constructed using the standard \RXTE\ data analysis
procedures. Subsequent spectral analysis then used the \textsc{xspec}\ 
package distributed by the High Energy Astrophysics Science Archive
Research Center (HEASARC).
\begin{table*}
\begin{minipage}{150mm}
\caption{Log of \XTE\ spectroscopic observations.}
\label{XTETable}
\begin{center}
\begin{tabular}{llc}
\hline
\noalign{\smallskip}
Date    & Start/End   & Duration \\
        &  (UT)       & (ks)     \\ 
\noalign{\smallskip}
\hline
\noalign{\smallskip}
May 14  & 00:07--00:28 & 1.26    \\
        & 01:42--02:04 & 1.32    \\
        & 03:24--03:40 & 0.96    \\
\noalign{\smallskip}
May 20  & 13:55--14:33 & 2.28    \\
        & 14:35--14:56 & 1.26    \\
        & 15:42--16:09 & 1.62    \\
        & 16:11--16:33 & 1.32    \\
        & 17:24--17:46 & 1.32    \\
        & 17:48--18:09 & 1.26    \\
\noalign{\smallskip}
May 27  & 17:29--18:19 & 3.00    \\
        & 19:07--19:55 & 2.88    \\
\noalign{\smallskip}
June 8  & 12:45--13:41 & 3.36    \\
        & 14:21--15:23 & 3.72    \\
        & 15:57--17:02 & 3.90    \\
        & 17:33--18:34 & 3.66    \\
\noalign{\smallskip}
June 20 & 12:55--13:12 & 1.02    \\
        & 14:31--14:54 & 1.38    \\
        & 16:07--16:35 & 1.68    \\
\noalign{\smallskip}
June 30 & 13:00-13:59 & 3.54     \\ 
\noalign{\smallskip}
\hline
\end{tabular}
\end{center}
\end{minipage}
\end{table*}
\subsection{\GRO\ data}
\label{CGROSection}
The BATSE Large Area Detectors (LADs), which were designed for the
study of gamma-ray bursts, have been used extensively and with great
success to monitor hard-X-ray sources using the earth-occultation
technique. Each of the eight LADs, located at the corners of the
\CGRO\ spacecraft in an octahedral geometry, are unshielded NaI
scintillators with a nominal $2\upi$\,sr field of view with
sensitivity over 16 energy channels covering about 20--1900\,keV. The
data are sampled at 2-second intervals (although the effective time
resolution here is $\sim1$ 90-minute spacecraft orbit). The useful
limiting sensitivity of the earth-occultation technique is typically
$\sim100$\, mCrab. For a complete description of this technique and
its capabilities see Harmon, et al.\ \shortcite{H92}.  We constructed
a light curve, spanning the interval covered by our \HST\ and \RXTE\
campaign, using the standard earth-occultation data products obtained
under the auspices of a \CGRO\ Guest Investigator Program designed to
study X-ray novae outbursts. The fluxes are derived from summations of
typically several days of data within a given viewing period, using
the weighted (by viewing angle) average of typically 2--3 LADs. Count
rates were converted to photon flux by assuming in this case a photon
power-law index, $\Gamma$, of 2.8 (where $N(E)\propto E^{-\Gamma}$)
and applying an absolute calibration. Energy channels covering
approximately 20--200\,keV were included in the analysis.  The
resulting light curve is shown in Figure \ref{LongLCFig}.  The
vertical axis is equivalent to photons\,cm$^2$\,s$^{-1}$, with the
$0.3\simeq1$\,Crab as indicated.  A nominal threshold is
0.03\,photons\,cm$^2$\,s$^{-1}$.
%
%
\section{Long-term multi-wavelength evolution}
\begin{figure}
\begin{center}
\epsfig{width=3in,height=4.5in,file=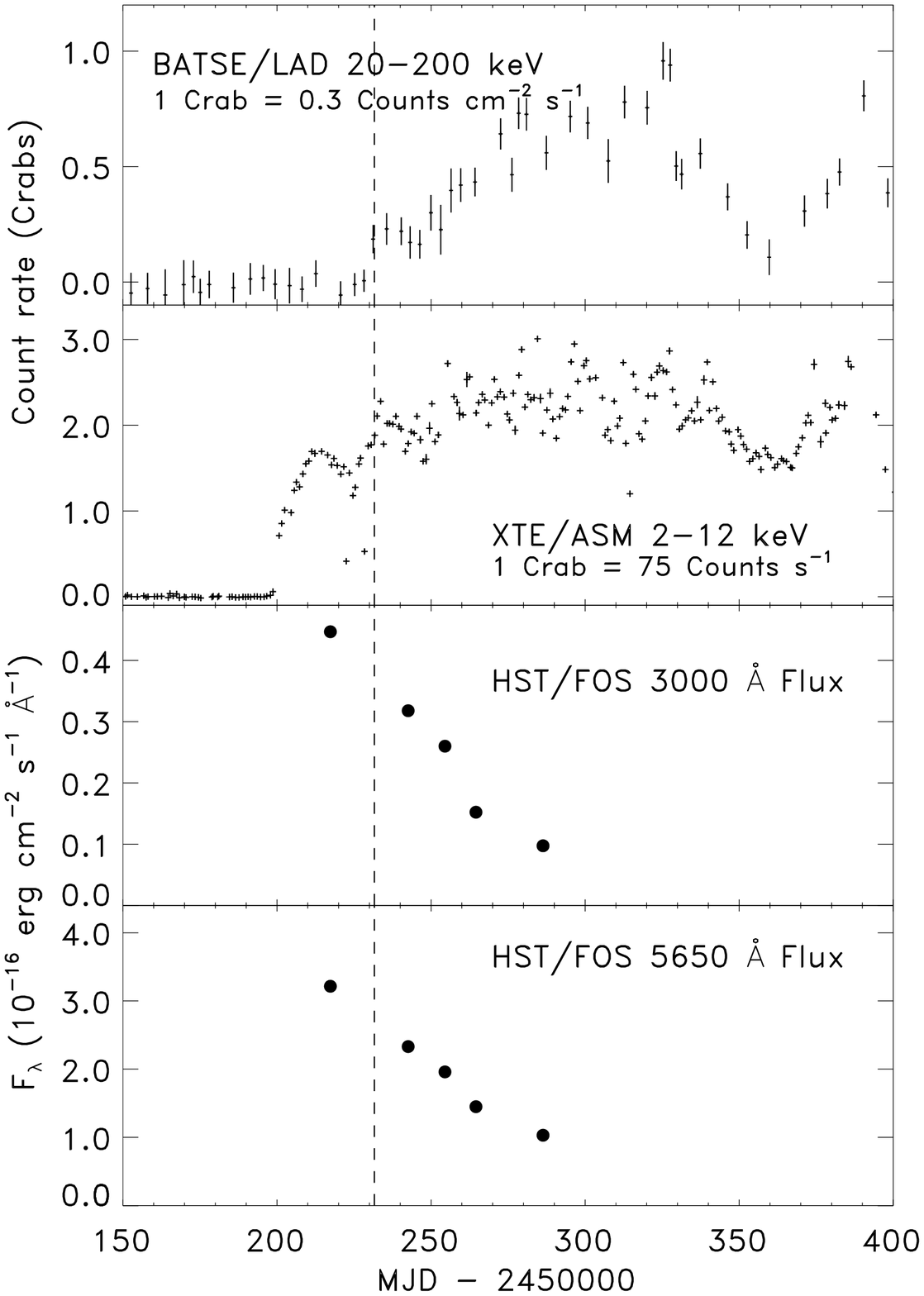}
\caption{Long term light curves of the 1996 outburst.  Statistical errors in
  the UV and optical points are comparable to or smaller than the
  symbols used.  The time-axis begins at 1996 March 8.  The discrepant
  behaviour of the optical-UV and X-ray data is clear.  The dotted
  line shows the first radio detection of this outburst.  The closest
  non-detection was 9~days before this, making the exact beginning of
  radio flare difficult to pinpoint.}
\label{LongLCFig}
\end{center}
\end{figure}

The general nature of the outburst is characterised by
Fig.~\ref{LongLCFig}, showing the \GRO\ BATSE (20--200\,keV), \XTE\
ASM (2--12\,keV) and \HST\ FOS light curves from Spring/Summer 1996.
Most striking is the contrast between the optical/UV decline and the
slow hard X-ray rise, an apparent anti-correlation.  We note that
Motch, Ilovaisky and Chevalier \shortcite{MIC85} observed a hard--soft
transition in the black hole candidate GX~339--4 during 1981 June in
which the {\em soft} X-ray rise was accompanied by an apparently
anticorrelated optical decline.  The timescales and spectral changes
involved in that transition were quite different to those in \novasco\
however, and there is not an obvious connection between the two cases.

Also notable is the difference between the rapid rise in the soft
X-ray band, and the delayed, slower rise in the hard X-ray band. This
is contrary to at least one other case, \novamus\, where the hard flux
rise preceded the soft \cite{E94}.  That outburst, however, exhibited
a canonical fast-rise, exponential decay light curve \cite{CSL97}
which was certainly not the case here.  Hameury et al.\
\shortcite{H97} have successfully modelled both the optical and soft
X-ray rise in \novasco\ using a two-component model involving a
conventional outer disc and an advective inner disc.  In this model,
the outburst is triggered by the limit cycle instability acting in the
outer disc, but the 6\, day soft X-ray delay is due to the need to
fill up the inner disc.

Radio detections of this outburst were short-lived; there certainly
was not the sustained radio activity, nor the large amplitudes, of the
1994 outburst, and no jets were detected.  Hjellming \& Rupen
\shortcite{HR96} report no detection above 0.5\,mJy between 1996
January and May~20.  On May~29 they detected \novasco\ at 19\,mJy
(4.8\,GHz) when Hunstead \& Campbell-Wilson \shortcite{HCW96} also
detected it at a flux of $55\pm5$\,mJy (843\,MHz).  This flare decayed
initially with an e-folding time 1.4\,days \cite{HWCW97} making it
much shorter lived than the activity at other wavelengths.  The first
detection of this flare is marked in Fig.~\ref{LongLCFig} and appears
to coincide with the initial step-like rise in the BATSE light curve.
The actual beginning of the flare could have been up to 9~days before
this however.  This contrasts with the outburst of 1994 when radio
flares {\em followed} hard X-ray outbursts.  There also appears to
have been an increase in soft X-ray variability at this time; this may
be related to the rise of the hard component.
%
%
\section{Interstellar extinction}
\subsection{Ultraviolet extinction}
\label{SeatonSection}
We use our ultraviolet spectra through the 2175\,\AA\ interstellar
absorption feature to estimate the reddening.  This method assumes
that the observed spectra can be fit by a reddened power-law, and has
previously been applied to the sources \novamus\ \cite{C92} and
\novaper\ \cite{S94}.  While the underlying spectrum may not be an
exact power-law, this should be a reasonable approximation over the
restricted spectral range in which the extinction curve varies
rapidly.  No other assumptions about the properties of \novasco\ are
required, although we do need to assume an extinction curve.  We adopt
the mean galactic extinction curve of Seaton \shortcite{S79}.  The
$\Rv$ dependent curves of Cardelli, Clayton and Mathis
\shortcite{CCM89} do yield similar values for the reddening if we
assume the average $\Rv=3.2$.  We also considered the unusual
extinction curve of \sigmasco\ \cite{CH93}, which lies 17$^{\circ}$
from \novasco.  The fit with this curve was significantly poorer than
with the S79 curve.

Initial attempts to fit a reddened $f_{\nu}\propto\nu^{1/3}$ power-law
(see Sect.~\ref{BBDiscSection}) to the whole spectrum \cite{H96}
suggested a value of $\EBV=1.3$.  This yielded a poor fit to the
2175\,\AA\ feature.  Examining the dereddened spectra however, we find
an intrinsic break around 2600\,\AA.  Analysis of background counts
and scattered light in the UV cannot explain this purely as a
calibration problem.  As this break appears in spectra dereddened with
a wide range of parameters, for all extinction curves considered, it
is unlikely to be an artefact of the dereddening process.  We
therefore base our revised reddening estimate on the UV data obtained
with the G160L grating only.  While both the G270H and PRISMs overlap
into the region below 2600\,\AA, this region is very sensitive to an
incorrect background subtraction (more so than the G160L in the
overlap region) and so including this data would be more likely to
systematically bias our reddening estimate than to improve it.

\begin{figure}
\begin{center}
\epsfig{angle=90,width=3in,height=2in,file=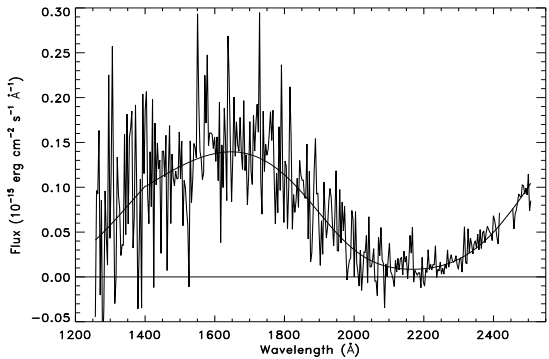}
\caption{Best fitting reddened power-law to 14 May data assuming all the 
  excess to be scattered light.}
\label{DerFitFig}
\end{center}
\end{figure}

\begin{figure*}
\begin{center}
\epsfig{angle=90,width=6in,height=4in,file=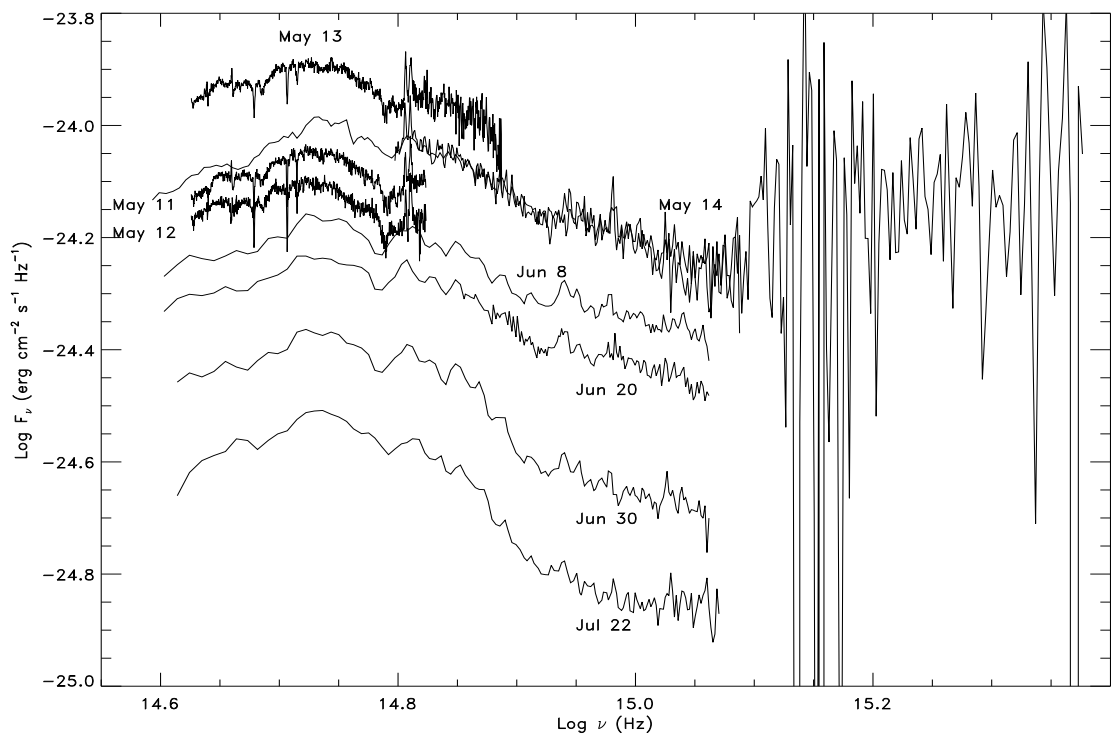}
\caption{Spectra dereddened using the S79 extinction law with $\EBV=1.2$.}
\label{SeatonDerFig}
\end{center}
\end{figure*}

We fit reddened power-law models to the average G160L spectra by
$\chi^{2}$ minimisation, using a robust grid-search method to locate
the minimum.  Errors are estimated from $1\sigma$ two-parameter
confidence intervals as defined by Lampton, Margon \& Bowyer
\shortcite{LMB76}.  Our estimates are shown in
Table~\ref{SeatonTable}.  They are not sensitive to whether the excess
light is taken to be scattered light or particle induced but they are
strongly affected by the {\em level} subtracted.  The fit to the
spectrum assuming all of the excess to be scattered light, shown in
Fig.~\ref{DerFitFig}, is extremely good.  The dereddened spectra are
shown in Fig.~\ref{SeatonDerFig}.
\begin{table}
\caption{Best fits to May 14 UV observations using a power-law
  spectrum reddened with the galactic average extinction curve.  We
  compare various models for the subtraction of excess light from the
  UV spectrum in order to assess the systematic uncertainty this
  introduces.  See Fig.~\ref{G160CountsFig} for an illustration of the
  highest and lowest background estimates.}

\label{SeatonTable}
\begin{center}
\begin{tabular}{lccc}
\noalign{\smallskip}
\hline
\noalign{\smallskip}
Background model              & $\EBV$          & $\alpha$      & $\chi^2_R$ \\
\noalign{\smallskip}
\hline
\noalign{\smallskip}
Pure scattered light,         &                 &               &            \\
\hspace*{2em}best estimate    & $1.19\pm0.07$   & $0.35\pm0.16$ & 1.10       \\
\noalign{\smallskip}
Pure particle background,     &                 &               &            \\
\hspace*{2em}best estimate    & $1.21\pm0.07$   & $0.47\pm0.17$ & 1.10       \\
\hspace*{2em}lowest estimate  & $1.08\pm0.06$   & $0.57\pm0.15$ & 1.10       \\
\hspace*{2em}highest estimate & $1.34\pm0.09$   & $0.37\pm0.19$ & 1.12       \\
\noalign{\smallskip}
\hline
\end{tabular}
\end{center}
\end{table}
\subsection{Direct optical estimation}
We can also obtain a direct estimate of the visual extinction, $\Av$,
by considering the expected absolute magnitude of the secondary star.
Orosz \& Bailyn \shortcite{OB97} obtain a best fit spectral type of
F5\,IV, with an acceptable range of F3--F6\,IV and an effective radius
of $4.85\pm0.08$\,R$_{\sun}$.  Using absolute magnitudes and radii of
main sequence F3--F6 stars \cite{G92} we can rescale the magnitudes to
the effective radius of \novasco\ thereby approximating the absolute
magnitude of the secondary star to be $M_{\rm V}=0.7\pm0.5$.  Orosz \&
Bailyn \shortcite{OB97} observe a mean magnitude in quiescence of
$m_{\rm V}=17.12$ and estimate by fitting photospheric absorption
lines that the disc contributes $<10$\, per cent of the visible light.
Assuming a disc contribution of $\sim5$\, per cent gives a mean
apparent magnitude of the secondary star of $17.18\pm0.06$.  The
errors in this figure are significantly less than in our estimate of
the absolute magnitude.  Adopting a distance of $3.2\pm0.2$\,kpc (see
Sect.~\ref{ParameterSection}), we obtain $\Av=4.0\pm0.5$ hence,
assuming an average extinction curve ($\Rv=3.2$), $\EBV=1.25\pm0.17$.
This estimate is dependent on the assumed system parameters, and so
vulnerable to systematic errors in these.
\subsection{Interstellar absorption lines}
In keeping with the highly reddened nature of \novasco, we see a rich
spectrum of interstellar features in our \AAT\ spectra. We measure the
equivalent width of the unresolved Na~D doublet to be
$2.5\pm0.1$\,\AA\ and also identify diffuse interstellar bands
\cite{He95} at 5778/80, 6177 and 6284\,\AA.  Bianchini et al.
\shortcite{B97} make more precise measurements of the equivalent
widths of the Na~D doublet (5890\,\AA) and the 6613\,\AA\ interstellar
band of 2.26\,\AA\ and 0.27\,\AA\ respectively.  They deduce $\EBV =
1.30$ and 0.97 respectively.

This method should be viewed with caution however.  Firstly, Bailyn et
al.\ \shortcite{B95} obtained a much larger Na~D equivalent width of
4.5\,\AA\ at a time when the doublet was blended with \HeI\ emission.
In spite of this they obtain a {\em lower} reddening of $\EBV=1.15$
simply by using a different equivalent width--reddening relation.
Secondly, as noted by Munari and Zwitter \shortcite{MZ97}, the Na
D-lines are not in general sensitive to $\EBV \geq 0.5$ due to
saturation, so this method can underestimate the reddening; whether it
does or not depends on the detailed substructure of the lines which is
not usually resolved.

There may also be signs of the very broad structure (VBS) that has
previously been reported in extinction curves of highly reddened
objects \cite{KMS86}.  This consists of very shallow dips in the
extinction curve between 5000\,\AA\ and 8000\,\AA\ and shows up as
apparent weak, broad emission features centred at 5300, 5900 and
6800\,\AA.  We can possibly identify VBS in our spectra, in particular
around 6800\,\AA, (see Fig.~\ref{CompSpecFig}).
\subsection{X-ray absorption}
\label{XRayAbsSection}
The most precise measurements of the absorption column density are
those obtained by fitting \ASCA\ spectra.  Inoue et al.\
\shortcite{I94} derive $N_{\rm H}=5\times10^{21}$\,cm$^{-2}$ from
observations on 1994 August~23.  Subsequent observations obtained
$4.4\times10^{21}$\,cm$^{-2}$ \cite{N94} and
$8\times10^{21}$\,cm$^{-2}$ \cite{I95}.  Greiner, Predahl \& Pohl
\shortcite{GPP95} use the \ROSAT\ high-resolution imager to resolve
the dust scattering halo.  Modelling this they deduce $N_{\rm
H}=7\times10^{22}$\,cm$^{-2}$.
\subsection{Summary}
It is encouraging that direct estimates of the visual extinction,
measurements of interstellar absorption lines and fitting the
2175\,\AA\ feature all give consistent results.  We adopt the \HST\
value as our best estimate of the reddening: $\EBV=1.2\pm0.1$.  This
can be compared with the extinction maps of the Galactic plane
presented by Neckel \& Klare \shortcite{NK80}.  \novasco\ lies close
to the boundary of two of their regions, one of low extinction, for
which we would expect $\EBV \sim 0.4$ at 3.2\,kpc, and one of high
extinction, for which the highest extinctions measured (up to
$\sim$2.5\,kpc) correspond to $\EBV = 1.3$.  Our measurement is thus
consistent with the position of \novasco.

Furthermore, using the gas to dust scaling of Bohlin et al.\
\shortcite{BSD78}, and their estimate of the scatter in this
relationship, this would correspond to a range of hydrogen column
densities of $N_{\rm H}=4.6-10\times10^{21}$\,cm$^{-2}$, in good
agreement with the range of \ASCA\ values observed.  In subsequent
analysis we adopt the intermediate \ASCA\ measurement of $N_{\rm
H}=5\times10^{21}$\,cm$^{-2}$, but allow this to vary between
$3-8\times10^{21}$.  We ignore the \ROSAT\ result as it disagrees
with \ASCA, optical and UV measurements.
%
%
\section{Spectral lines}
\begin{figure*}
{\bf a)}\epsfig{angle=90,width=3in,height=2in,file=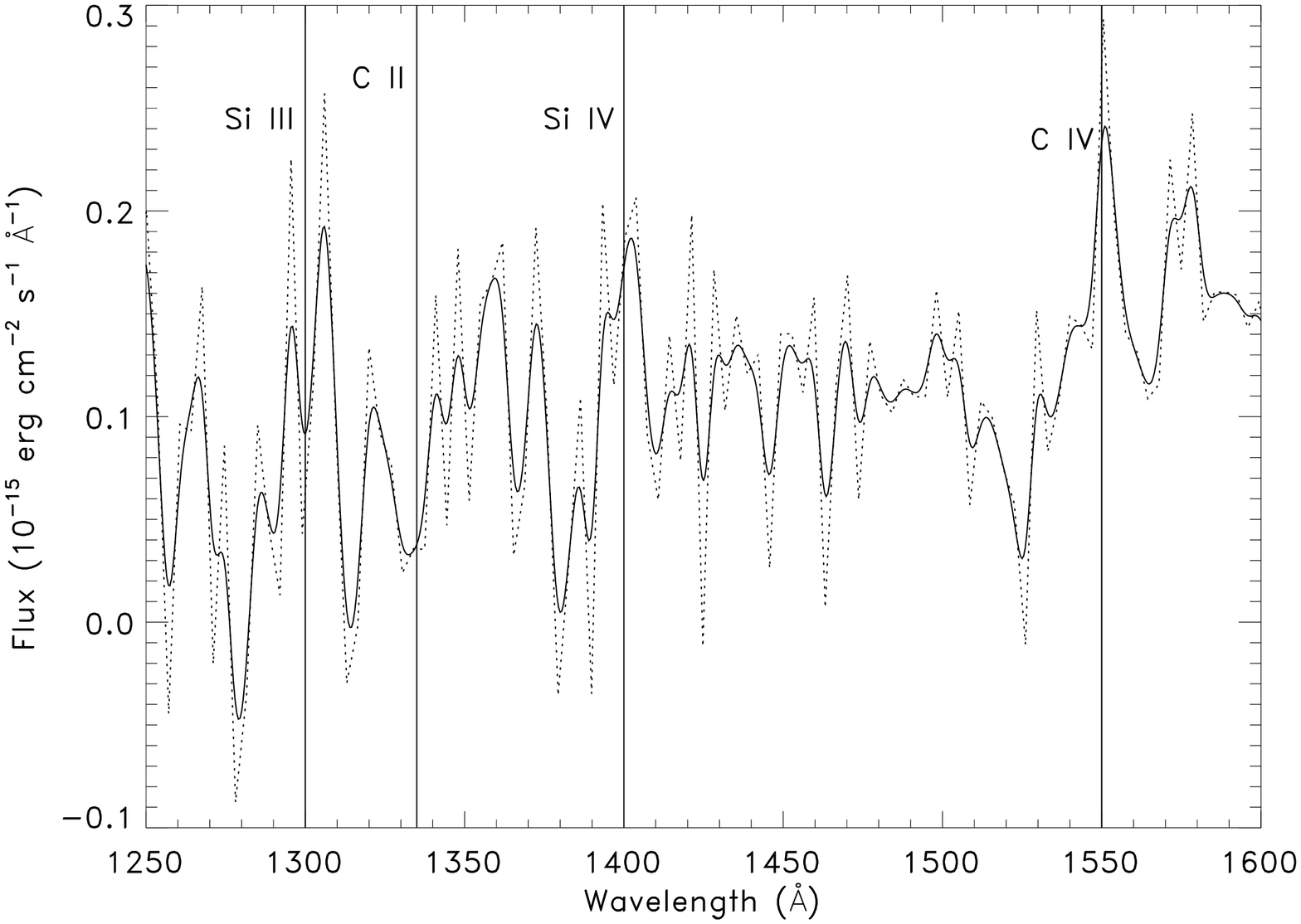}
{\bf b)}\epsfig{angle=90,width=3in,height=2in,file=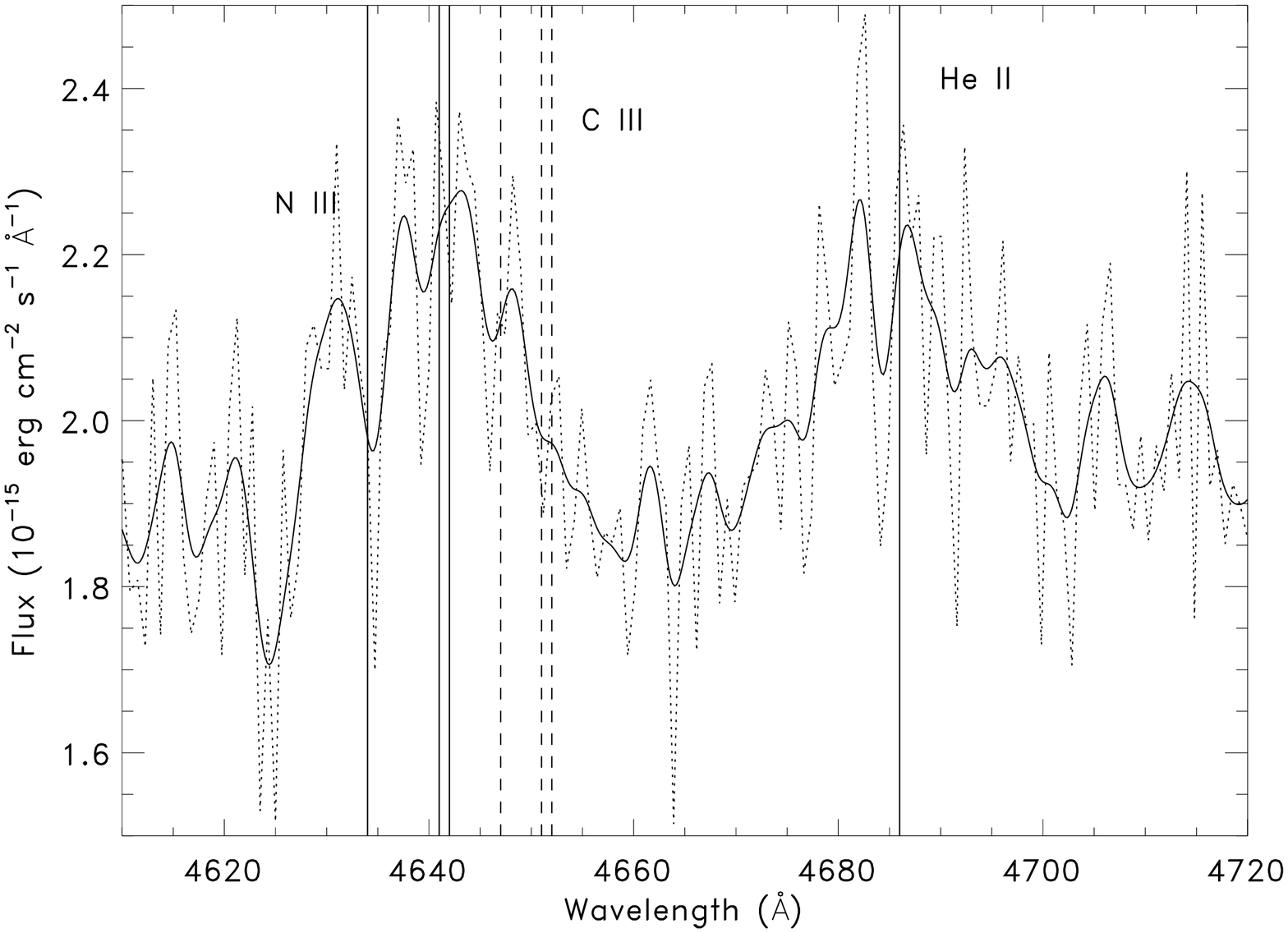}
{\bf c)}\epsfig{angle=90,width=3in,height=2in,file=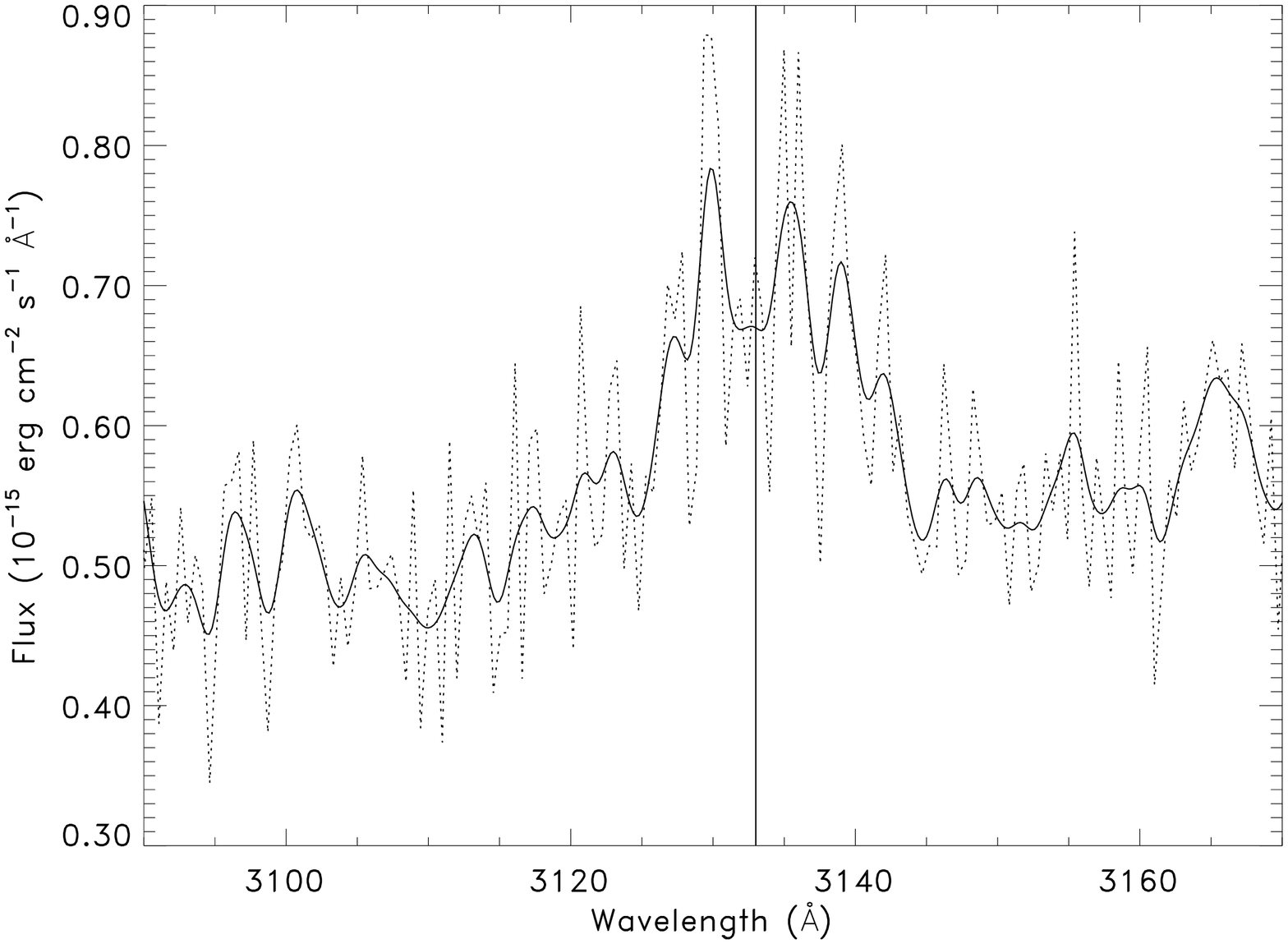}
{\bf d)}\epsfig{angle=90,width=3in,height=2in,file=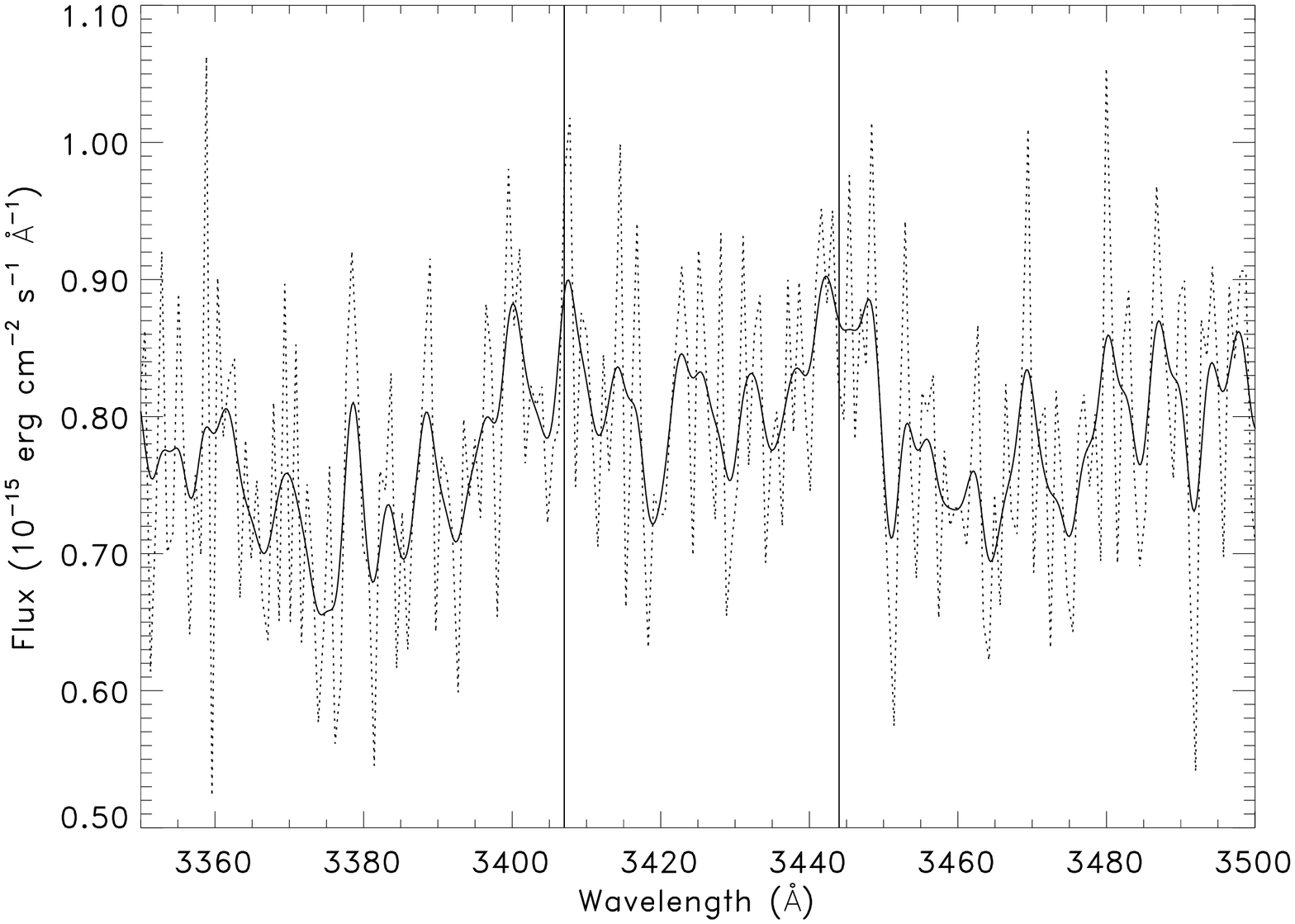}
\caption{\HST\ line profiles.  The dotted line shows the raw data; the
  solid line is the same smoothed with a width equal to a resolution
  element.  a) Possible P~Cygni profiles in the UV resonance lines.
  b) The \NIII\ \lamlam4634,\,4641,\,4642 and \HeII\ \lam4686 emission
  lines.  The positions of the \CIII\ \lamlam4647,\,4651,\,4652 lines
  are marked with dotted lines.  c) The \OIII\ \lam3133 emission line.
  d) The possible \OIII\ \lamlam3407,\,3444 emission lines.}
\label{LineFig}
\end{figure*}
\begin{table*}
\begin{minipage}{120mm}
\caption{Properties of emission lines detected in May spectra.}
\label{EmissionTable}
\begin{center}
\begin{tabular}{llcccc}
\noalign{\smallskip}
\hline
\noalign{\smallskip}
Line & Telescope/ & Fitted     & Line flux                              & 
EW    & FWHM  \\
     & date       & wavelength & ($10^{-15}$\,erg\,s$^{-1}$\,cm$^{-2}$) & 
(\AA) & (\AA) \\
\noalign{\smallskip}
\hline
\noalign{\smallskip}
\OIII (\lam3133)              & \HST/May 14 & $3133.2\pm0.7$ & $2.9\pm0.4$ & 
$5.6\pm0.8$ & $14\pm2$                                               \\
\noalign{\smallskip}
\OIII (\lam3407)              & \HST/May 14 & $3405.0\pm2.6$ & $2.6\pm0.9$ & 
$3.5\pm1.3$   & $21\pm5$                                             \\
\noalign{\smallskip}
\OIII (\lam3444)              & \HST/May 14 & $3439.3\pm4.3$ & $2.3\pm0.9$ & 
$3.1\pm1.4$   & $21\pm5$                                             \\
\noalign{\smallskip}
\NIII (\lamlam4634,\,41,\,42) & \AAT/May 11 & $4638.9\pm0.9$ & $7.7\pm0.5$ & 
$5.2\pm0.3$ & $26\pm2$                                               \\
                              & \AAT/May 12 & $4637.1\pm1.1$ & $7.3\pm0.8$ & 
$5.7\pm0.7$ & $24\pm3$                                               \\
                              & \AAT/May 13 & $4635.7\pm0.9$ & $11.0\pm1.0$ & 
$5.2\pm0.5$ & $27\pm2$                                               \\
                              & \HST/May 14 & $4640.8\pm0.8$ & $6.3\pm0.6$ & 
$3.4\pm0.4$ & $16\pm2$                                               \\
\noalign{\smallskip}
\HeII (\lam4686)              & \AAT/May 11 & $4685.0\pm0.9$ & $4.6\pm0.5$ & 
$3.0\pm0.3$ & $18\pm3$                                               \\
                              & \AAT/May 12 & $4686.2\pm1.1$ & $2.7\pm0.6$ & 
$2.0\pm0.5$ & $11\pm3$                                               \\
                              & \AAT/May 13 & $4681.7\pm0.7$ & $8.5\pm0.9$ & 
$4.0\pm0.4$ & $18\pm2$                                               \\
                              & \HST/May 14 & $4686.1\pm0.9$ & $6.2\pm0.6$ & 
$3.3\pm0.3$ & $21\pm2$                                               \\
\noalign{\smallskip}
\hline
\end{tabular}
\end{center}
\end{minipage}
\end{table*}
\subsection{Ultraviolet resonance lines}
We see likely P~Cygni profiles in the UV resonance lines: \SiIII\
(\lam1302), \SiIV\ (\lam1397) and \CIV\ (\lam1549) together with
possible \CII\ (\lam1335) absorption: Fig.~\ref{LineFig}a.  Although
the detection of these is marginal, their positions are correct and
the peak to trough separation of all three P~Cygni profiles
corresponds to velocities of order 5000\, km\,s$^{-1}$; comparable to
(although somewhat higher than) those seen in dwarf novae in outburst
(Shlosman \& Vitello 1993 and references therein).  Shlosman \&
Vitello model the expected disc wind profiles in the \CIV\ 1550 line
accounting for the non-radial outflow and solving the radiative
transfer problem in detail.  They find that for low inclination
systems where the disc is seen through the wind, only a blue-shifted
absorption component is seen.  For high inclination systems, symmetric
emission is seen from the wind.  Only for a relatively narrow range of
inclinations do we expect to see `classical' P~Cygni profiles showing
absorption and emission in similar strengths.  This range of
inclinations is around 60--70$^{\circ}$, in striking agreement with
the inclination determined for \novasco\ in quiescence, which adds
support to this interpretation.  The noise level of our observations
prevents distinguishing between such a disc wind and a pure radial
outflow, possibly emanating in the inner regions.
\subsection{Helium and Bowen fluorescence lines}
The only helium line detected in the spectrum is \HeII\ P$\alpha$
(\lam4686), illustrated in Fig.~\ref{LineFig}b.  There is no detection
of \HeII\ P$\beta$ (\lam3203) or \HeII\ B$\alpha$ (\lam1640).

We detect three \OIII\ Bowen lines, representing both the O1
(\lamlam3133, 3444) and O3 (\lam3407) cascades (see below).  These are
shown in Figs.~\ref{LineFig}c and \ref{LineFig}d.  Only the line at
\lam3133 is unambiguously detected; this is theoretically expected to
be the strongest line of the O1 channel.

There is also emission around 4641\,\AA\ (Fig.~\ref{LineFig}b) which
we interpret as primarily due to a blend of three \NIII\ Bowen lines
(\lamlam4634,\,41,\,42).  It is possible that there is some
contribution from the non-Bowen \CIII\ lines (\lamlam4647,\,51,\,52).
Our data are of insufficient quality to deblend these lines.

The Bowen fluorescence mechanism as it applies in X-ray binaries has
been well discussed from a theoretical standpoint by Deguchi
\shortcite{D85} and in the context of observations of Sco~X-1 by
Schachter, Filippenko \& Kahn \shortcite{SFK89}.  The essence of the
mechanism is this: the \HeII\ Ly$\alpha$ transition (\lam303.783) is
nearly coincident with two transitions of \OIII.  \HeII\ photons can
excite these and the subsequent decays, through the O1 and O3 channels
respectively, produce UV emission lines of \OIII.  One of these decay
transitions, the O4 channel, is in turn nearly coincident with a
doublet of \NIII.  This can then be excited and decay to produce
\NIII\ emission lines.  Using the measured line fluxes of the
principal related lines we estimate the efficiency of this process, as
measured by the Bowen yields.

The oxygen yield is the probability that a \HeII\ Ly$\alpha$ photon
will produce a cascade through the O1 channel.  It is measured by
\begin{equation}y_{\rm HeO1}=k_{\rm H}\frac{f(\lambda 3133)}{f(\lambda 4686)}\end{equation}
where f($\lambda$) is the \textit{dereddened} line flux.  Following
Schachter et al. \shortcite{SFK89}, we adopt $k_{\rm H}=0.28$.  This
value is dependent on the temperature of the producing region, but the
uncertainty this introduces is comparable to the uncertainty in the
line flux measurements and much less than that in dereddening the flux
ratio.  Using the Seaton \shortcite{S79} extinction curve with
$\EBV=1.2\pm0.1$, we obtain
\begin{equation}y_{\rm HeO1}=0.70\pm0.15.\end{equation}
Deguchi \shortcite{D85} predicts values for this yield between 0.5 and
0.8; Schachter et al. \shortcite{SFK89} find values ranging from
0.47--0.59.  Our results are consistent with these, and are not
precise enough to further constrain the emission line region.

In an analogous way, the nitrogen yield, $y_{\rm ON}$, represents the
fraction of O4 photons which produce a nitrogen cascade.  The
situation is complicated here by two factors.  Firstly we are unable
to deblend a possible \CIII\ component from the \NIII\ blend, so we
really only have an upper limit on the \NIII\ flux.  Secondly, to
perform this calculation correctly we need to know the strengths of
both the O1 and O3 cascades, since both can in turn initiate an O4
cascade.  We are unable to reliably measure any O3 lines so we cannot
quantify this contribution.  If we simply ignore the O3 contribution
and define
\begin{equation}y_{\rm ON}=k_{\rm KM}\frac{f(\lambda\lambda 4634,\,41,\,42)}
                            {f(\lambda 3133)}\end{equation}
where $k_{\rm KM}=8.6$, independent of physical conditions, then this will also
lead to us overestimating the yield.  We determine
\begin{equation}y_{\rm ON}\leq3.7\pm0.8\end{equation}
Again, this is in reasonable agreement with Schachter et al.
\shortcite{SFK89} who find $y_{\rm ON}=3.2-4.0$ for Sco~X-1, with the
same assumptions; they also note that this is an overestimate.
\subsection{A constraint on the extreme ultraviolet flux}
\label{EUVSection}
We can also use the observed line flux of the \HeII\ \lam4686 line to
place an upper limit on the EUV flux.  Photoionisation in the disc
and/or secondary star by 55--280\,eV photons (bounded by the
ionisation energy of \HeII\ and the carbon K-edge) will produce
\HeIII, which will then recombine leading to \lam4686 emission.  We
can estimate the EUV flux which will give the observed line emission.
This is a very useful constraint, as this region of the spectrum can
never be observed directly for a source as highly reddened as
\novasco.  Since \lam4686 could also originate in collisional
excitation at the stream impact point, this will be an upper limit.
It is also very dependent on the assumed geometry, both of the EUV and
line emitting regions.  The method has been developed in the context
of cataclysmic variables, where the EUV flux is believed to originate
at the boundary layer of the disc and white dwarf (Patterson \&
Raymond 1986, Marsh \& Horne 1990).  Marsh, Robinson and Wood
\shortcite{MRW94} demonstrated that it could also be applied to SXTs,
with the EUV instead originating from the inner regions of the disc.
Their calculation assumes that the EUV radiation is isotropic,
appropriate for a point source or a corona.  If it originates in the
optically thick inner accretion disc then because both the secondary
star and the outer disc see these regions at a high inclination, this
will substantially overestimate the EUV flux incident upon them.  We
account for this inclination effect by making a more rigorous
assessment of the fraction of EUV luminosity intercepted in both disc
and secondary star cases.

This inclination effect introduces a factor of $\cos\theta$ into the
EUV angular flux distribution, where $\theta$ is the angle of the EUV
radiation to the disc normal.  We assume that there is no shielding by
the inner part of the disc and consider two cases: radiation
intercepted by the disc only and radiation intercepted by the
secondary star assuming a vanishingly thin disc.

In the disc case the fraction, $\alpha$, of EUV luminosity intercepted
is
\begin{equation}\alpha \simeq \left(\frac{H}{R}\right)^{2}\end{equation}

We treat the secondary star case by modelling it as a circle of radius
$r_{eq}$ centred on the secondary position.  $r_{eq}$ can be
determined numerically (as a function of the mass ratio, $q$) by requiring
that the circle subtend the same solid angle at the primary as the
real Roche lobe does.  For \novasco\ ($q=2.99$), $r_{eq}=0.286$,
somewhat larger than the polar radius of 0.269.  The fraction of flux
intercepted is then
\begin{equation}\alpha \simeq \frac{2r_{eq}^3}{3\upi}\end{equation}

We represent the \lam4686 line luminosity by $L_{4686}$ and the
integrated 55--280\,eV luminosity by $L_{\rm EUV}$.  A fraction
$\alpha$ of the total number of 55--280\,eV photons are intercepted by
the disc or secondary star and a fraction $\epsilon$ of the total
number of photoionisations caused by 55--280\,eV photons recombine
through this channel.  For Case B recombination (appropriate when the
emission line region is optically thick to the ionising radiation but
optically thin to the recombination lines), $\epsilon\simeq0.2$.  The
ratio of luminosities will be
\begin{equation}\frac{L_{\rm EUV}}{L_{4686}}=\frac{1}{\alpha\epsilon}
                                E_{\rm EUVE}
                                \left(\frac{\lambda_{4686}}{\rm hc}\right)\end{equation}

where $E_{\rm EUVE}$ is an average energy of the EUV photons.  The EUV
flux, $F_{\rm EUV}$, that we would observe is related to the total
luminosity by $F_{\rm EUV}=L_{\rm EUV}\cos i/2\upi d^2$, where $d$ is
the distance of the system.  For optically thin recombination lines,
the line emission will be isotropic.  The relation between flux and
luminosity then depends only on the fraction, $\beta$, of the emission
line region that we see: $F_{4686}=\beta L_{4686}/4\upi d^2$.  For a
disc, we see one face only and $\beta=0.5$.  For lines originating on
the secondary star, $\beta$ is phase and inclination dependent.  The
ratio of EUV to line fluxes is

\begin{equation}
\frac{F_{\rm EUV}}{F_{4686}}=\frac{2\cos i}{\beta\alpha\epsilon}
                                E_{\rm EUVE}
                                \left(\frac{\lambda_{4686}}{\rm hc}\right)
\end{equation}

We observe line fluxes in the range
(2.7--8.5)$\times10^{-15}$\,erg\,s$^{-1}$\,cm$^{-2}$.  If these are
produced in the disc, then they should be phase independent, and we
must interpret this range as due to stochastic variability.  We then
should use the line flux from May 14 to obtain the best estimate of
the May 14 EUV flux.  Dereddening this (assuming $\EBV=1.2\pm0.1$) and
taking $\epsilon=0.2$, $E_{\rm EUVE}=100$\,eV, an EUV bandwidth of
55--280\,eV, $\lambda_{4686}=4686$\,\AA\ and assuming a minimum
fractional disc half thickness of $H/R=0.01$, as was taken by Marsh et
al. \shortcite{MRW94}, this translates into an {\em upper} limit on
the EUV flux per unit frequency interval of
$f_{\nu}(2.4\times10^{16}$\,Hz$)\la5.9\times10^{-24}$\,%
erg\,cm$^{-2}$\,s$^{-1}$\,Hz$^{-1}$ for the case where \lam4686
emission originates only in the disc.

Examining the four line fluxes as a function of orbital phase,
however, the flux is highest at spectroscopic phase 0.19, when we see
the heated face of the secondary star nearly face on, and lowest at
phase 0.81, when we see the rear face of the secondary star.  This
suggests that the \lam4686 emission originates at least partially on
the secondary star and so this cannot be completely shielded by the
disc.  With this interpretation we can use the highest \lam4686 flux,
when we see virtually all of the emission region ($\beta\sim1$) to
constrain the EUV flux to be
$f_{\nu}(2.4\times10^{16}$\,Hz$)\la2.3\times10^{-25}$\,%
erg\,cm$^{-2}$\,s$^{-1}$\,Hz$^{-1}$.

The latter calculation represents the more realistic limit, in the
sense that if the disc did become very thin then the \lam4686 flux
would not become vanishingly small; it would instead be produced on
the secondary star. We stress the caveat that we have assumed none of
the EUV flux is shielded by the inner disc.  If this assumption were
invalid then this EUV constraint becomes meaningless.
%
%
\section{Modelling the optical and ultraviolet spectra}
\label{ModellingSection}
We now discuss possible interpretations of the optical/UV spectra.  In
view of the large uncertainty in determining the intrinsic flux
distribution from such highly-reddened spectra, together with the
inherent complexity, it would not be appropriate to attempt to fit a
detailed model for the complete spectra.  Instead we will seek a
simple characterisation of the observed spectra and then compare this
with the available sources of radiation.

We obtain the best fitting parameters by $\chi^2$ minimisation.  Since
the scatter about the fit is predominantly due to intrinsic and
interstellar features rather than measurement uncertainties, we obtain
very large $\chi^2$ values and applying conventional $\chi^2$
confidence region methods leads to unrealistically small error
estimates.  In line with the simple characterisation approach we
therefore do not attempt to construct rigorous confidence limits.

\begin{figure}
\begin{center}
\epsfig{width=3in,height=4.5in,file=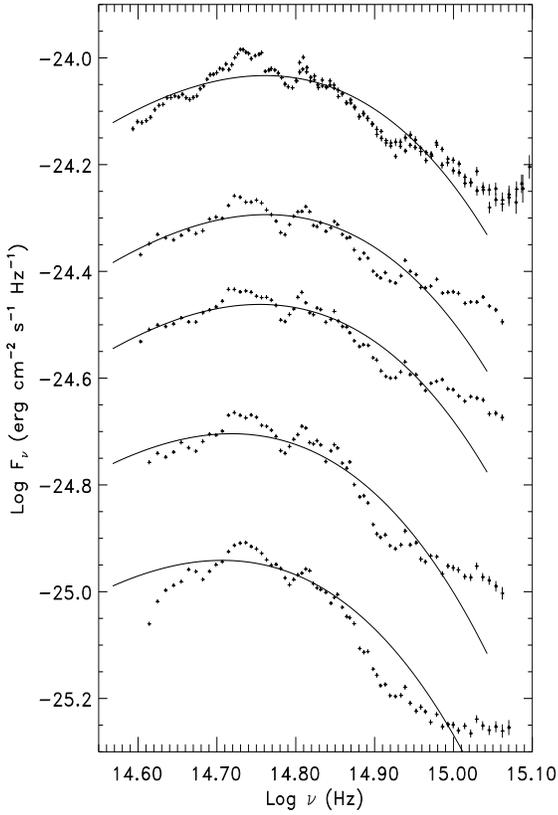}
\caption{Blackbody fits to the optical excess in the \HST\ spectra
  ($\lambda\ge2600$\,\AA).  The fits are poor in detail and are only
  intended to characterise the this coarse properties of this
  component.  In order to separate the successive visits clearly, a
  downward shift of 0.1 has been introduced in each visit relative to
  the one above it i.e.\ the lowest visit has been shifted downwards
  by 0.4.}
\label{BBFitFig}
\end{center}
\end{figure}

Since we have the most comprehensive wavelength coverage in the first
(May) \HST\ visit, we will concentrate on modelling that spectrum,
noting differences in the other epochs as they arise.  We immediately
notice that there appear to be two distinct components: a power-law
($\alpha\simeq0.33$) in the UV and a second component peaking in the
optical.  The power-law suggests an obvious interpretation as the
predicted spectrum of a steady state blackbody accretion disc.  The
optical component considered in isolation can be characterised by the
best-fitting blackbody.  The deduced parameters are shown in
Table~\ref{BBParameterTable} and the fits to the spectra in
Fig.~\ref{BBFitFig}.  This does not give a good fit in detail, but
will give us an approximate estimate of the properties of a thermal
model.  We note in passing that a blackbody does give a better fit to
observations than stellar spectra of similar temperatures; the
temperatures suggested by the spectra correspond to A~stars and hence
very large Balmer jumps which we do not see.
\begin{table}
\caption{Properties of blackbody fits to the optical and near UV 
  ($\lambda\geq2600$\,\AA) \HST\ spectra.}
\label{BBParameterTable}
\begin{center}
\begin{tabular}{lcc}
\noalign{\smallskip}
\hline
\noalign{\smallskip}
        & T(K)   & A($\times10^{23}$\,cm$^2$) \\
\noalign{\smallskip}
\hline
\noalign{\smallskip}
May  14 & 9800 & 5.0 \\ 
June  9 & 9900 & 3.4 \\
June 20 & 9700 & 3.1 \\
June 30 & 8900 & 2.9 \\
July 22 & 8700 & 2.2 \\
\noalign{\smallskip}
\hline
\end{tabular}
\end{center}
\end{table}

Of course to assume that the spectrum longwards of 2600\,\AA\ is
purely due to one component is not realistic.  The two components will
overlap, and whilst the exponential Wien tail of the optical component
could drop off rapidly in the UV, it is less clear that the power-law
component will drop off so rapidly in the optical.  This blackbody
characterisation thus represents an upper limit on the size of the
optical component.  The alternative extreme, to simply extrapolate the
power-law into the optical would result in approximately half the flux
at the peak of the optical spectrum coming from the power-law.  It
would not change the position of the peak of the optical component
dramatically, so this component should have roughly the same
temperature and no less than half the area given in
Table~\ref{BBParameterTable}.

We now move on to consider physical interpretations of these two
components in more detail.
\subsection{The steady state blackbody disc model}
\label{BBDiscSection}
Our first visit G160L data is suggestive of the $\nu^{1/3}$ spectrum
expected from a steady state optically thick accretion disc; a similar
UV spectral index was found in \HST\ observations of X-ray Nova Muscae
1991 \cite{C92}.  While a SXT in outburst is clearly not in a steady
state, we might expect this to be a good approximation on the decline
from outburst if the viscous timescale of the disc is shorter than the
timescale of mass transfer rate variations.  The spectrum in this case
is discussed by Frank, King and Raine \shortcite{FKR92} and Cheng et
al.\ \shortcite{C92}.

For frequencies, $\nu$, such that
\begin{equation}{\rm k}T(R_{\rm out}) \ll 
  {\rm h}\nu \ll {\rm k}T(R_{\rm in})\end{equation}
where $R_{in}$ and $R_{out}$ are the radii of the inner and outer
edges of the disc and h is Planck's constant, the spectrum reduces to
the often referred to $\nu^{1/3}$ disc spectrum:
\begin{equation}f_{\nu}=f_{0}\frac{\cos i}{d^{2}}\left(m\dot{m}\right)^{2/3}
  \nu_{15}^{1/3}\int_{0}^{\infty}\frac{x^{5/3}dx}{{\rm e}^{x}-1}\end{equation}
where $f_{0}\simeq 2.9\times
10^{-26}$\,ergs\,s$^{-1}$cm$^{-2}$\,Hz$^{-1}$, $i$ is the inclination,
$d$ is the distance in kpc, $m$ is the compact object mass in
$M_{\sun}$, $\dot{m}$ is the mass transfer rate in
$10^{-9}$\,M$_{\sun}$\,yr$^{-1}$, $\nu_{15}$ is the frequency in units
of $10^{15}$\,Hz and the integral evaluates to 1.9.

We can fit this spectrum to our data, and with the parameters of
Sect.\ \ref{ParameterSection} deduce a value of $\dot{M}\simeq8\times
10^{-7}$\,M$_{\sun}$\,yr$^{-1}$, adopting our best estimate of
background counts and reddening.  Allowing the full range of
backgrounds and reddening values discussed earlier gives a range of
$1\times 10^{-7}$\,M$_{\sun}$\,yr$^{-1}\leq \dot{M}\leq 7\times
10^{-6}$\,M$_{\sun}$\,yr$^{-1}$ We can compare this with the Eddington
accretion rate of $\dot{M}_{\rm Edd}=1.6\times
10^{-7}$\,M$_{\sun}$\,yr$^{-1}$ (taking a compact object mass of
7\,M$_{\sun}$ and assuming and accretion efficiency of 10 per cent.
The observed accretion rate (near the peak of the outburst) is thus of
the order of the Eddington rate, as might be expected intuitively.

The extrapolation of this spectrum into the extreme UV however would
predict $f_{\nu}\left(2.4\times10^{16}\,{\rm Hz}\right)\simeq1.7\times
10^{-24}\, {\rm erg\,s^{-1}\,cm^{-2}\,Hz^{-1}}$.  This exceeds our
constraint on the flux at this frequency (see Sect.\ \ref{EUVSection})
and suggests that either this extrapolation is invalid, or that the
assumptions used in deriving the EUV constraint, for example that the
inner disc does not obstruct the EUV flux, are inappropriate.  We will
return to this question in Sect.~\ref{OutburstModelSection}.
\begin{figure}
\begin{center}
\epsfig{angle=90,width=3in,height=2in,file=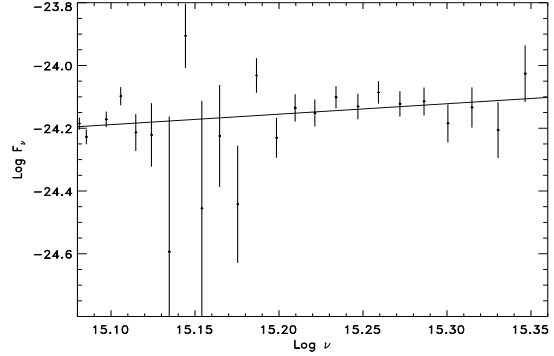}
\caption{Fit of a $\nu^{1/3}$ spectrum to the rebinned G160L data
  to determine \.{M}.}
\label{ChengFitFig}
\end{center}
\end{figure}
\subsection{The secondary star contribution}
\label{SecContrib}
While we could reasonably expect the UV spectrum to be dominated by an
accretion disc, in the optical we need to account for the secondary
star contribution as well.  We first assess the expected contribution
if the secondary star spectrum is unaffected by the outburst.  In
accordance with Orosz \& Bailyn \shortcite{OB97} we represent the
secondary star contribution using the spectrum of the F5\,IV star
BD\,+630013 \cite{GS83}.  Orosz \& Bailyn \shortcite{OB97} present
quiescent V band light-curves and estimate that the disc contributes
only 5 per cent of this light.  Using the phases from Table
\ref{ExpTable} and the extinction solution of Sect.
\ref{SeatonSection} we estimate {\em secondary star} V band dereddened
magnitudes and normalise our secondary star spectrum to this level to
obtain a best estimate of the quiescent secondary star contribution
for each \HST\ visit.  The phases and magnitudes are summarised in
Table \ref{SecMagTable}.
\begin{table}
\caption{Phases and deduced secondary star magnitudes for each \HST\ 
  visit.  Photometric phase zero is when the secondary is on the near side
  of the compact object.}
\label{SecMagTable}
\begin{center}
\begin{tabular}{lccc}
\noalign{\smallskip}
\hline
\noalign{\smallskip}
Date    & Photometric   & \multicolumn{2}{c}{Quiescent secondary star } \\
        & phase         & \multicolumn{2}{c}{magnitude (V)}        \\
\cline{3-4}
        &               & observed & dereddened                    \\
\noalign{\smallskip}
\hline
\noalign{\smallskip}
May  14 & $0.67       $ & 17.08    & 13.24                         \\
June  9 & $0.38       $ & 17.15    & 13.31                         \\
June 20 & $0.94       $ & 17.28    & 13.44                         \\
June 30 & $0.75       $ & 16.98    & 13.14                         \\
July 22 & $0.05       $ & 17.29    & 13.45                         \\
\noalign{\smallskip}
\hline
\end{tabular}
\end{center}
\end{table}

\begin{figure}
\begin{center}
\epsfig{angle=90,width=3in,height=2in,file=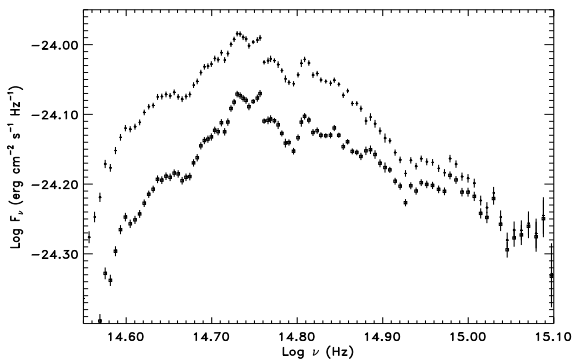}
\caption{The May 14 PRISM spectrum before and after subtracting an estimate
  of the {\em quiescent} spectrum at the same orbital phase.}
\label{SecSubFig}
\end{center}
\end{figure}

In Fig.~\ref{SecSubFig} we show the first visit PRISM spectrum before
and after subtracting this secondary star component.  The optical
excess is clearly not removed and so this cannot be accounted for by
including a non-irradiated secondary star contribution.

\begin{figure}
\begin{center}
\epsfig{angle=90,width=3in,height=2in,file=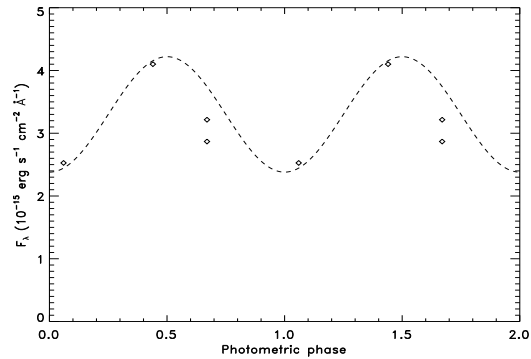}
\caption{Average 5600--5700\,\AA\ fluxes from the May \HST\ and \AAT\ 
  spectra plotted on photometric phase.  Statistical errors are
  comparable to or smaller than the symbols.  A sine wave with maximum
  at phase 0.5 has been overplotted to guide the eye; this is not
  intended to be a model or fit to the data.  These are suggestive of
  such a modulation, but not conclusive.}
\label{SpectralModFig}
\end{center}
\end{figure}

It is likely of course is that the secondary star is significantly
heated by irradiation during the outburst and that this increases its
spectral contribution.  If the optical spectrum were dominated by a
strongly irradiated secondary star then we would expect to see an
orbital modulation in the light curve, with a maximum at photometric
phase~0.5.  Figure~\ref{SpectralModFig} shows the average flux from a
representative continuum region (5600--5700\,\AA) from each May
spectrum plotted on the photometric phase.  These data sample the
light curve poorly, but do suggest a maximum at phase 0.5.  This would
be consistent with significant heating of the secondary, but is not
conclusive.

Quantitatively an irradiated secondary star model is plausible; the
estimated radius of 4.85\,R$_{\sun}$ \cite{OB97} corresponds to an average
area $\sim3.6\times10^{23}$\,cm$^2$, although this will be only an
upper limit on the visible, irradiated area.  Comparison with
Table~\ref{BBParameterTable} shows that this is a little small, but at
least of the right order of magnitude to explain the optical
component.  We find, however, that the optical excess is seen in all
spectra, with a shape largely independent of orbital phase, even for
phases for which the irradiated face of the secondary star will not be
visible.  We must conclude that an irradiated secondary star cannot
explain {\em all} of the excess, but it could provide the
phase-dependent part of the solution.
\subsection{A warm outer disc?}
We next consider if some or all of the excess could be coming from the
outer regions of the accretion disc.  While we have already invoked
the disc to explain the UV spectrum, this only requires the inner
regions of the disc; in that model most ($>80$ per cent) of the disc
is too cool to make a significant contribution to the UV.  We could
therefore remove most of the disc, {\em or give it a completely
different temperature distribution}, without destroying our model for
the UV spectrum.  Two possible ways in which the outer disc
temperature distribution are likely to be changed will be considered
below.  First, however, let us estimate the available area.  Adopting
a disc radius of 80 per cent of the effective Roche lobe radius
\cite{E83} and an inclination of $\sim$70\,$^{\circ}$, we estimate a
projected area of $\sim2.1\times10^{23}$\,cm$^2$.  In addition, we may
have to include radiation from the outside edge of the disc since
\novasco\ is a high-inclination system.  De Jong, van Paradijs and
Augusteijn \shortcite{dJvPA96} have found that for some persistent low
mass X-ray binaries, the disc half-thickness, $H/R$ can be greater
than 0.2.  With the inclination of \novasco\ and a disc of this
thickness, the projected area of the edge amounts to $\sim$70 per cent
of the projected area of the disc face, increasing the total area
presented by the disc to $\sim3.6\times10^{23}$\,cm$^2$, comparable to
that of the secondary star.  As is the case with the secondary, this
is of the right order of magnitude to explain the observed radiation.
\label{IrradiationSection}

If thermal emission from the outer disc is responsible for the
observed optical component, this would require us to heat these
regions up to $\sim$10\,000\,K.  Is this possible?  Qualitatively,
this is what we would expect in the presence of a significant
irradiating flux.  The exact results depend sensitively upon the shape
of the concave disc surface, but in general a disc in which energy
generation is dominated by irradiation is expected to have a flatter
temperature distribution than a steady state disc dominated by viscous
dissipation. For example, Vrtilek et al.\ \shortcite{V90} derived
$T\propto R^{-3/7}$ for the irradiation dominated limit instead of
$T\propto R^{-3/4}$.  When both dissipation and irradiation are
considered, the inner disc will be dominated by dissipation whilst
irradiation will raise the temperature of the outer disc above that of
a steady state dissipative model.  The spectrum expected is then
qualitatively similar to that observed except that there is a very
smooth spectral transition between the two regimes (irradiation
dominated optical emission and dissipative far-UV) in contrast to the
sharp break we observe.  It may be that using more realistic local
spectra than black bodies and including the effects of limb darkening
will explain the sharpness of the break.  Limb darkening, in
particular, is very strong in the UV \cite{DWH96}, is both temperature
and wavelength dependent and can be expected to be significant for a
high inclination disc.

\label{HighStateSection}

Even in the absence of irradiation, there is another mechanism which
will cause the temperature distribution in the outer disc to deviate
from that of a steady state disc.  This is the thermal-viscous limit
cycle instability \cite{C93}, in which for a given annulus of the
disc, there is a range of effective temperatures for which the annulus
is both thermally and viscously unstable.  We would not expect any of
the disc to lie within this range.  The exact temperatures involved
are dependent on the disc properties and detailed physics of hydrogen
ionisation, but of order 6\,000--10\,000\,K.  For most SXTs this is
not a problem as it is possible, for plausible mass transfer rates, to
have a steady state disc in outburst which is entirely hotter than the
unstable temperature range.  \novasco\, however, in common with
V404~Cyg is a long period system, which means it has a much larger
disc than many other SXTs.  Because of this, even for a steady state
disc accreting at the Eddington rate, part of the outer disc would be
cool enough that it should lie within the unstable temperature range:
in this Eddington limit case, a temperature of 10\,000\,K is reached
at 0.23 of the primary lobe radius.  The implication is then that for
\novasco, there may exist {\em no globally stable steady state
solution with a sub-Eddington accretion rate}.  The steady state model
{\em must} therefore break down in the outer disc.  What will happen
is less clear.  The most obvious expectation would be that outside the
radius at which the disc is just within the hot stable state there
will be an abrupt transition to the cool state.  This would imply that
most of the disc can never participate in the outburst.  Can this
really be the case?  An alternative that would be consistent with our
data is that instead of the outer disc falling into the cool state, it
is maintained in the hot state.  This would put most of the visible
disc (and plausibly also the disc outside edge) at temperatures of
$\sim$10\,000\,K\ and could {\em quite naturally} explain why the
optical component has this temperature and maintains it throughout the
outburst.
\subsection{Self-absorbed synchrotron emission}
Finally we suggest an intriguing alternative to thermal models.  Could
the cool excess actually be self-absorbed synchrotron emission?  This
model has more parameters and can give a better fit to the observed
spectrum than the blackbody model, in particular for the rapid
turnover at the peak.  Can such a model work quantitatively?

We adopt the simplest possible model for producing this spectrum.  We
assume a cloud thickness $l$, area $l^2$, volume $l^3$.  This cloud
contains a uniform magnetic field $B$ and a power-law distribution of
relativistic electrons, $n_{\rm e}(E)=\kappa E^{-p}$.  For simplicity,
we assume equipartition between the magnetic fields and electrons.
The expected spectrum is derived by Longair \shortcite{L94} and we do
not reproduce the details.  The functional form of the spectrum is:
\begin{equation}I_{\nu}=A\nu^{5/2}\left[1 - \exp(-B\nu^{(-p+4)/2})\right]\end{equation}
Our problem is then to fit $A$, $B$ and $p$ and then to determine the
underlying physical parameters.

\begin{figure}
\begin{center}
\epsfig{width=3in,height=4.5in,file=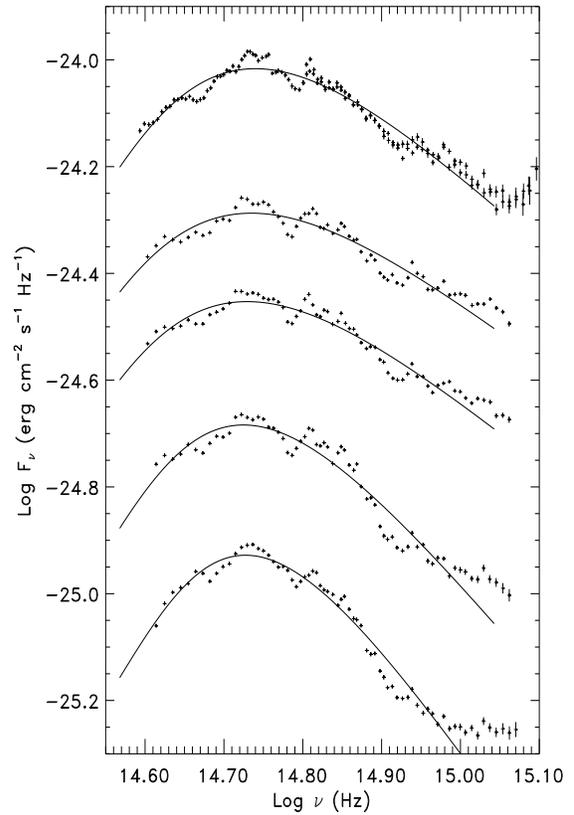}
\caption{Self-absorbed synchrotron fits to the cool optical excess in
  the \HST\ spectra ($\lambda\ge2600$\,\AA).  A downward shift has
  been introduced as in Fig.~\ref{BBFitFig} to separate the successive
  visits.}
\label{SynchrotronFig}
\end{center}
\end{figure}

The best fits with such models are shown in Fig.~\ref{SynchrotronFig}
and the deduced parameters are given in Table~\ref{SynchrotronTable}.
We immediately note that the source must be very compact.  This is
actually the best constrained of the parameters as for a self-absorbed
synchrotron source, the observed flux depends only on the angular size
of the source and the magnetic field.  With an estimate of the
distance (3.2\,kpc) the derived linear size then scales with only the
fourth root of the assumed magnetic field.

The clearest evolution in the spectra is the overall decline in flux
level; this translates mainly into a shrinking of the inferred
synchrotron source.  There is also a steepening of the spectrum.  This
may be due to a change in the electron power-law index, as modelled,
or it may be due to a cut-off in the electron distribution above which
inverse-Compton and synchrotron losses dominate.  Such a cut-off could
also contribute to the sharpness of the spectral turnover near
2600\,\AA.

Synchrotron emission is often considered exclusively a radio
phenomenon.  That this is not the case is demonstrated by the
existence of optical counterparts to extragalactic radio jets.
Observations of the other galactic superluminal jet source
GRS~1915+105 have, revealed infrared jets, with the same position
angle as the radio jets \cite{SES96} and infrared flares with
strikingly similar durations, timescales and energies to those seen in
the radio at a similar time \cite{F97}.  Both of these results are
suggestive of infrared synchrotron emission.  It has also been
suggested that the optical spectrum of the SXT V404~Cyg in quiescence
may be dominated by self-absorbed synchrotron emission from an
advection dominated accretion flow \cite{NBM97}, and that the same is
true for \novasco\ \cite{H97}.  Our suggestion is thus speculative but
not completely without precedent.

\begin{table}
\caption{Parameters of self-absorbed synchrotron models for \HST\
  optical and UV data.  The Schwarzschild radius of a 7\,M$_{\sun}$
  black hole corresponds to $2.1\times10^6$\,cm.  In each case,
  deduced electron energies are $\gamma\sim$50--90,estimated from the
  observed synchrotron frequency range of $14.6\leq\log\nu\leq15.0$.
  Electron densities are lower limits obtained by integrating the
  electron distribution over this energy range.}
\label{SynchrotronTable}
\begin{center}
\begin{tabular}{lcccc}
\noalign{\smallskip}
\hline
\noalign{\smallskip}
        & $p$ & $l$             & $B$  & $N_{\rm e}$              \\
        &     & (R$_{\rm sch}$) & (kG) & ($10^{12}{\rm cm}^{-3}$) \\
\noalign{\smallskip}
\hline
\noalign{\smallskip}
May  14 & 3.7 & 97              & 54   & 7.6                      \\ 
June  9 & 3.3 & 80              & 61   & 8.3                      \\
June 20 & 3.4 & 76              & 59   & 8.4                      \\
June 30 & 4.5 & 70              & 48   & 7.8                      \\
July 22 & 5.3 & 60              & 42   & 7.6                      \\
\noalign{\smallskip}
\hline
\end{tabular}
\end{center}
\end{table}
%
%
\section{Modelling the X-ray spectra}
We now move on to consider our X-ray spectra.  We constructed
composite high-energy spectra covering the $\sim$2--200\,keV region
using summed data from the \RXTE\ PCA and the BATSE/LAD
earth-occultation as described in Sect.~\ref{CGROSection}.  We were
advised by the \RXTE\ PCA instrument scientists that the detailed
detector calibrations were still evolving, but nonetheless, reasonable
fits to the data were obtained in a number of cases, and some basic
spectral characteristics of the event could be revealed. We found that
for a variety of fitting scenarios, the lowest energy channels
($\la4$\,keV) proved to be problematic in that they contributed
significantly to the overall $\chi^2$.  Given the extreme high count
rates in these channels, $\chi^2$ values could become untenable very
readily, exceeding 100 in some cases. We additionally found channels
49 and 51 to be problematic in some cases, presumably a result of the
background estimates used, and we ignored these in our fitting as
well.

The general character of the high-energy spectra is that of a
power-law with a soft thermal excess component--i.e., the
high-soft-state characteristic of black hole candidate X-ray sources
accreting at super-Eddington rates.  The soft component could be
fitted by a multi-colour accretion disc, or equally well by a single
temperature blackbody.

\begin{figure}
\begin{center}
\epsfig{angle=90,width=3in,height=2in,file=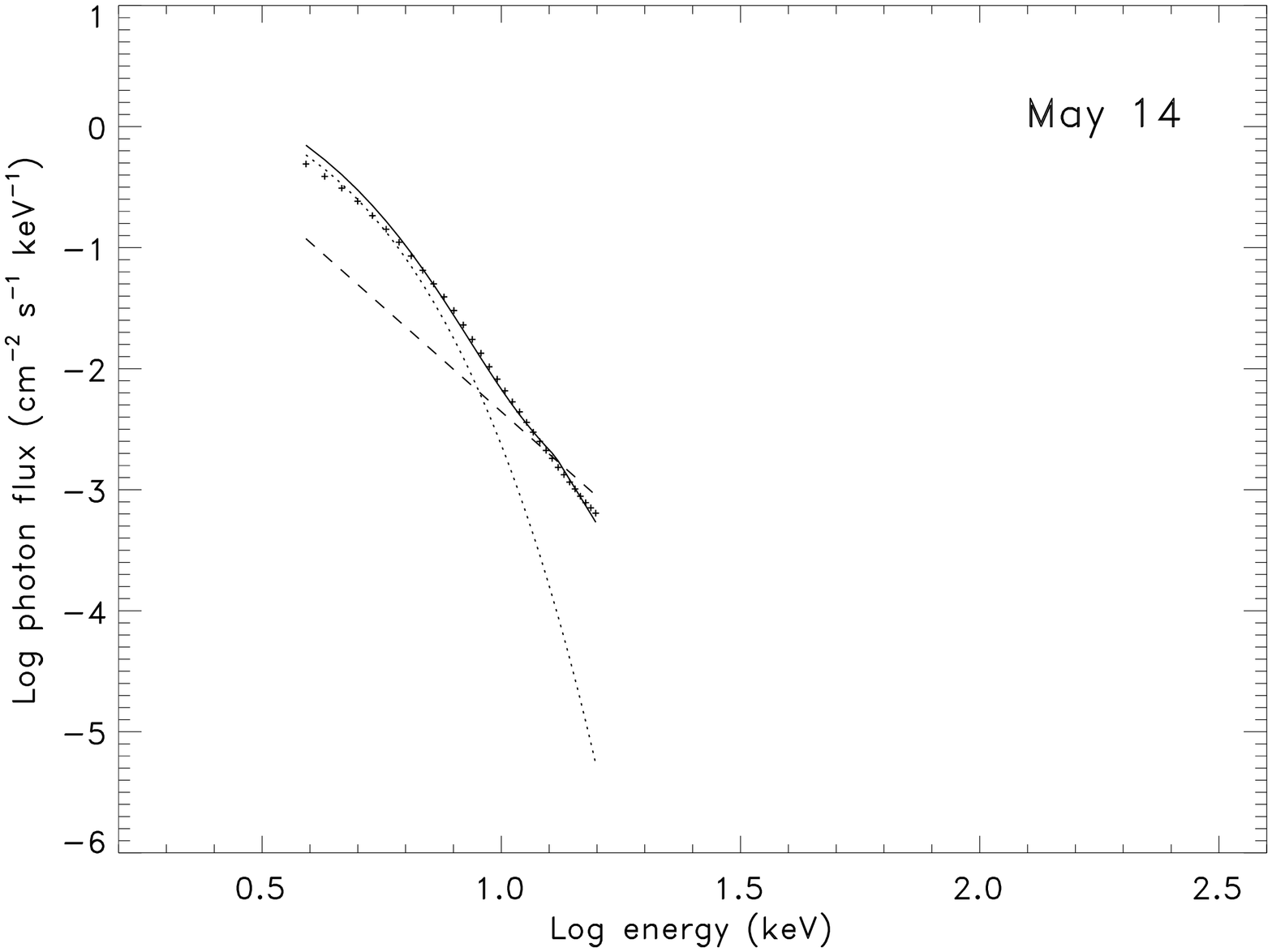}
\epsfig{angle=90,width=3in,height=2in,file=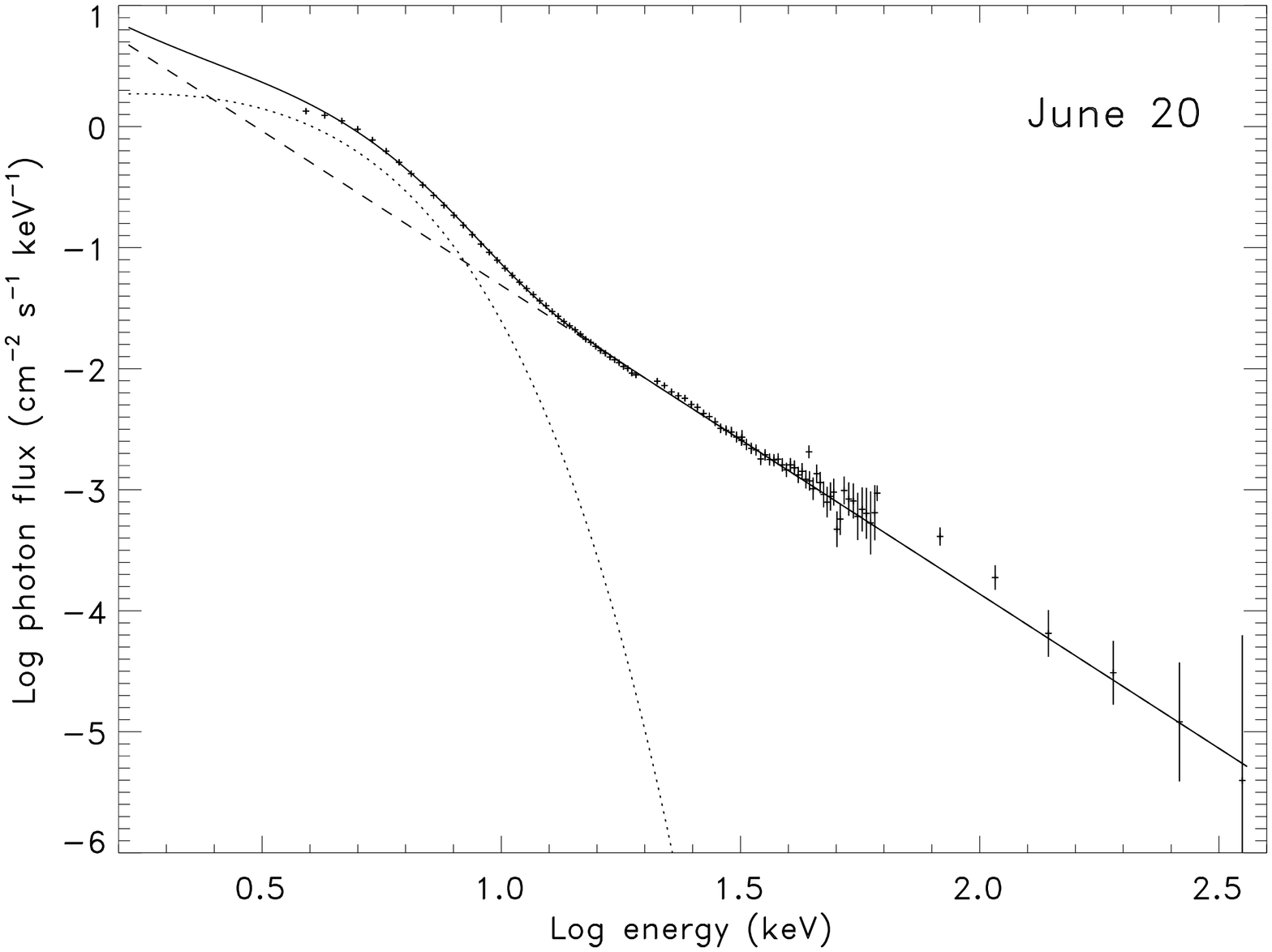}
\caption{Two component fits to the \CGRO\ BATSE and \RXTE\ PCA
  spectra from 1996 May 14 (upper panel) and June 20 (lower panel).
  The solid line shows the overall fit to the spectrum.  The dashed
  and dotted lines respectively show the power-law and thermal
  components.  In the May 14 spectrum an exponential cut-off has been
  included in the {\em overall} fit as described in the text.}
\label{XRayJun20Fig}
\end{center}
\end{figure}

Prior to the high-energy `turn-on', i.e., when the source surpassed
the BATSE threshold, the PCA spectra required a soft power law (photon
index $\Gamma\simeq3.5$) plus a blackbody with characteristic
temperature $T\simeq0.8$\,keV, and an exponential cut-off (e-folding
scale of 5.4\,keV) above $\simeq13$\,keV.  The fit to the May~14
spectrum is illustrated in Fig.~\ref{XRayJun20Fig}.  The background
subtracted PCA flux was consistent with zero above $\sim20$\,keV,
consistent with the non-detection by BATSE. The typical $\chi^2$ value
obtained were unsatisfactory for the purpose of attributing any one
specific model to the data.

At later epochs, although the overall PCA count rates were similar to
the initial epochs, the spectrum was harder and the BATSE/LADs could
now accumulate sufficient counts to constrain the high energy spectral
distribution. By June~20, the joint PCA/BATSE spectrum was modelled as
a power-law with photon index $\Gamma\simeq2.5$ and a ${\rm
  k}T\simeq1.1$\,keV blackbody. No high-energy cut-off was required,
the power-law now extended to $\ga100$\,keV above which the data
become noise limited.  In this case, the total reduced $\chi^2$ values
obtained were much more reasonable, $\sim$2 per degree of freedom. The
fit to the June~20 spectrum is illustrated in Fig.~\ref{XRayJun20Fig}.
It would thus appear that for this outburst, \novasco\ exhibited two
distinct substates of the canonical high-soft state characteristic of
black hole candidates.
\begin{table}
\caption{Parameters of fits to May 14 and June 20 X-ray spectra.}
\label{X-rayFitTable}
\begin{center}
\begin{tabular}{llc}
\noalign{\smallskip}
\hline
\noalign{\smallskip}
May 14  & Photon index, $\Gamma$      & $3.50\pm0.05$                 \\
        & Temperature, kT             & $0.825\pm0.001$\,keV             \\
        & Column density, N$_{\rm H}$ & $3.5\times10^{21}$\,cm$^{-2}$ \\
        & Cut-off energy              & $13.0\pm0.1$\,keV             \\
        & e-fold energy of cut-off    & $5.4\pm0.2$\,keV              \\ 
        & $\chi^2_{\rm R}$            & 7.168                         \\
\noalign{\smallskip}
June 20 & Photon index, $\Gamma$      & $2.55\pm0.02$                 \\
        & Temperature, kT             & $1.09\pm0.02$\,keV            \\
        & Column density, N$_{\rm H}$ & $8.0\times10^{21}$\,cm$^{-2}$ \\
        & $\chi^2_{\rm R}$            & 2.088                         \\
\noalign{\smallskip}
\hline
\end{tabular}
\end{center}
\end{table}
%
%
\section{Towards a consistent multi-wavelength model for the outburst}
\label{OutburstModelSection}
\begin{figure}
\begin{center}
\epsfig{angle=90,width=3in,height=2in,file=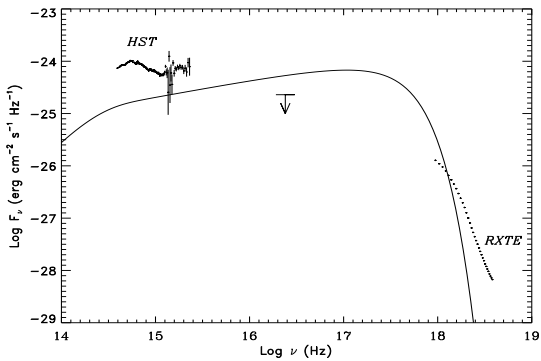}
\caption{Composite spectrum for the May~14 observations.  Shown are the
  \HST\ spectrum (left), the EUV constraint derived from the \HeII\ 
  \lam4686 flux (centre; see Sect.~\ref{EUVSection}) and the \XTE\ 
  spectrum (right).  Overlaid is a simplistic model spectrum for a
  steady state accretion disc accreting at the Eddington rate.  This
  assumes an outer radius of 80 per cent of the Roche lobe and an
  inner radius of three Schwarzschild radii.  No attempt has been made
  to fit this to the actual data and it is shown only to indicate the
  simplest expectation for the spectral energy distribution.}
\label{MWFig}
\end{center}
\end{figure}
Thus far we have dealt with each energy band and observation largely
in isolation.  In Fig.~\ref{MWFig} we show our composite May 14
spectrum.  For comparison we have overplotted the spectrum of a steady
state disc extending from three Schwarzschild radii to 80 per cent of
the Roche lobe.  This is a very simplistic model: black body local
spectra have been used throughout and no attempt has been made to
allow for limb darkening or relativistic effects.  The excess seen in
the optical can, as described earlier, be attributed to an outer disc
heated to hotter than it would be in the steady state model; X-ray
irradiation could be causing this.  The far UV spectrum could be
consistent with a steady state model but the higher energy data are
clearly not: the EUV constraint and X-ray spectrum both fall
significantly below an extrapolation of the far UV power-law.  This
makes our interpretation of the far UV spectrum questionable.  While
we would not expect the X-ray power-law tail to fit this model, the
thermal component is usually attributed to the inner disc and so
should represent the X-ray spectrum of the disc.  An explanation for
the power-law tail (a ubiquitous feature of black hole candidates) has
recently been suggested by Titarchuk, Mastichiadis \& Kylafis
\shortcite{TMK96}.  They propose that it is produced by {\em bulk
  motion} Comptonisation from infalling material very close to the
black hole event horizon.  Their models, whilst still preliminary,
predict a power-law spectrum extending to high energies, with the
observed spectral index.

Moving on to consider the evolution of the outburst, our most
challenging result for such a model to explain is the apparent
anti-correlation between optical/UV and hard X-ray behaviour.  In the
conventional picture of SXT outbursts driven by the limit cycle
instability we would expect a fast rise--exponential decay pattern at
both optical and X-ray energies (Chen et al. 1997).  If reprocessed
X-rays dominate the optical energy budget then the problem is even
worse, as we would expect the optical output to track the X-ray input
closely.  How can these difficulties be overcome?

The discrepant X-ray behaviour may be a natural consequence of the
long period of \novasco; most other CVs and SXTs have much shorter
periods and smaller discs (although we note that the X-ray transient,
V404 Cyg, has a longer period of 6.5\,d, but showed a an overall X-ray
decline, albeit with large superposed variability \cite{Oo97}).  The
longer period could lead to a longer timescale for the disc to settle
into a quasi-steady state and hence a continuing transfer of mass into
the inner regions giving the sustained high X-ray fluxes.  Detailed
modelling of how the limit cycle instability would operate in such a
system remains to be done.  This may explain why the X-ray activity
does not decline as expected.  The inverse problem -- that irradiation
by a high X-ray flux should keep the optical fluxes high -- could
possibly be explained geometrically.  Chen, Livio \& Gehrels
\shortcite{CLG93} considered the secondary shielding the disc from
irradiation in their model of the secondary maxima (reflares).  It is
also true, however, that the effectiveness of irradiation of the disc
is greater for a thicker disc as it will intercept more X-ray flux.
The shape of the disc is also crucial: we will only have direct
irradiation of the disc for a concave surface.  Finally the X-ray
albedo is likely to be higher if the X-rays fall on the disc at a
steep angle of incidence and this will tend to further amplify the
dependence on disc thickness and shape.  So if the shape and/or
thickness of the disc evolves significantly during the outburst, the
optical flux could still drop, in spite of the irradiating flux
increasing.

As noted in Sect.~\ref{HighStateSection}, the observed optical
spectral evolution could very naturally be explained if a large
fraction of the disc is maintained just in the high state, possibly by
irradiation.  Consider the following scenario.  At the beginning of
the outburst a heating wave propagates through the disc.  If the disc
then thins or changes its shape in such a way that the X-ray energy
input to a given annulus is decreasing, then a cooling wave will be
able to move inwards at the point at which the disc can no longer be
maintained in the high state.  This would very naturally produce the
fixed temperature--shrinking area behaviour inferred.  The limit-cycle
instability then acts as a thermostat ensuring that the outer edge of
the disc stays at a roughly constant temperature.  Our observation of
this fixed temperature evolution may therefore be direct evidence that
the instability is at work.

It is also worth noting an important difference between this outburst
and those previous.  Both Bailyn et al. \shortcite{B95} in the 1994
outburst and Bianchini et al.  \shortcite{B97} during 1995 observed
strong Balmer line emission.  In contrast we see very little Balmer
emission, and broad, shallow, absorption troughs dominate.  Bianchini
et al. infer from their H$\alpha$ light curves and the changes in the
width of the line during the outburst that it was produced by a strong
burst of mass transfer at the beginning of the outburst, triggered by
the {\em preceding} hard X-ray rise.  That we see no strong Balmer
emission in 1996 (when there was only a very slow hard X-ray rise) is
consistent with this interpretation and is further evidence that this
outburst was qualitatively different to the previous ones.  This seems
to have been a pure disc outburst with no extra mass transfer from the
secondary.

Our alternative suggestion, in which the optical emission is dominated
by non-thermal synchrotron radiation has intriguing possibilities.
This model could explain why the optical flux seems to be
anti-correlated with the hard X-ray flux (which we believe to be {\em
non-thermal} in origin) but appears completely unrelated to the {\em
thermal} soft X-ray flux.  The details of such a model are beyond the
scope of this paper however.
%
%
\section{Conclusions}
We have obtained a series of co-ordinated optical, UV and X-ray
spectra spanning several months of the outburst of an SXT.  Although
the optical light curve shows the expected decline, it has a spectrum
different to that expected on theoretical grounds.  Conversely,
although the X-ray spectra showed the familiar high state form, the
X-ray fluxes continued to rise through the optical decline, contrary
to expectations.  We have considered various interpretations of the
observations and suggest the following possible scenario:

The outburst was triggered by a heating wave in the disc, causing an
optical rise as the outer disc enters the hot state and a soft X-ray
rise subsequently when the material starts to reach the inner disc.
Inflow to the inner regions continues to rise for some time as the
disc tries to find a steady state.  About a month after the initial
rise the accretion mode near the black hole changes.  This results in
a rise in the hard X-ray activity and the X-ray variability as the
extended hard power-law component becomes prominent.  This change is
accompanied by a brief radio flare.  While the X-ray activity is still
rising, the disc itself is changing in thickness and/or geometry so
that irradiation becomes less efficient allowing the cooling wave to
move inwards producing the drop in optical flux at a nearly fixed
temperature.  There is also some irradiation of the secondary star,
producing an orbital variation in the continuum fluxes and \HeII\
4686\,\AA\ emission.

There clearly remain important unanswered questions not only about
\novasco, but about the outbursts of SXTs in general.  There are many
theoretical avenues to be explored in seeking an explanation of these
observations, especially in the modelling of long period, large disc
systems and the exploration of non-thermal models for the optical
emission.

This work also has many useful lessons for the observer.  We have
demonstrated the value of co-ordinated, multi-wavelength campaigns in
ruling out interpretations which might be suggested by a part of the
dataset, but are inconsistent with the whole.  We suggest the
following priorities for future observations of SXT outbursts:

\begin{enumerate}
\item UV observations are {\em crucial} to such a campaign for the
  following reasons.  a) It is only in this region that we may be
  seeing the expected $\nu^{1/3}$ characteristic accretion disc
  spectrum in this dataset; identification of this is an important
  indicator of the disc temperature distribution.  b) It is in the UV
  resonance lines that we see the signature of an accretion disc wind.
  Higher resolution, higher signal to noise observations (possible
  only for a less extremely reddened source) will test models of
  accretion disc winds and allow an estimate of the mass loss rate.
  c) The 2175\,\AA\ interstellar absorption feature is our best tool
  in estimating the reddening of these typically highly reddened
  objects; other measures such as Na~D-lines are not always reliable
  in these cases.  Without a good estimate of the interstellar
  reddening we cannot determine and hence interpret the intrinsic
  spectrum.
\item The campaign should include spectra, or at least multi-colour
  photometry, spanning and adequately sampling several consecutive
  orbits.  This will allow us to separate orbital spectral modulations
  from random variability and distinguish between emission from an
  irradiated secondary star and the accretion disc.  In the event of
  the discovery of an unambiguously eclipsing SXT these observations
  would be of central importance, allowing eclipse mapping of the
  emission regions.
\item Comprehensive X-ray spectra spanning as wide an energy range as
  possible should be an integral part of the campaign.  We observe the
  high energy side of a thermal component, but lower energy coverage
  is required to accurately characterise this component and
  distinguish between a single temperature blackbody and a
  multi-colour disc.
\item The campaign should include good red and near infrared coverage
  to obtain improved characterisation of the long wavelength turnover
  in the spectrum.  This will help in discriminating between thermal
  and non-thermal emission, which show different long-wavelength
  limits, and in the thermal case will provide more comprehensive
  information on the cooler parts of the system.
\end{enumerate}
%
%
\section*{Acknowledgements}
RIH is supported by a PPARC Research Studentship.  Support for this
work was provided by NASA through grant numbers NAG5-3311 and
GO-6017-01-94A from the Space Telescope Science Institute, which is
operated by the Association of Universities for Research in Astronomy,
Incorporated, under NASA contract NAS5-26555 and also through grant
NAS5-32490 for the \RXTE\ project.  This work made use of the \RXTE\
and \CGRO\ Science Centers at the NASA Goddard Space Flight Center and
used the NASA Astrophysics Data System Abstract Service.  We
acknowledge Alastair Allan's assistance in the \AAT\ observations.
Thanks also to Jerome Orosz and Roberto Soria for kindly providing
their spectra of \novasco\ for comparison.  Thanks to Jeff Hayes, Tony
Keyes, Tony Roman and Michael Rosa at STScI for support.  RIH would
like to thank John Cannizzo, Phil Charles, Andrew King, Kandu
Subramanian and many others for stimulating discussions, and
Jean-Pierre Lasota for helpful comments on the manuscript.
%
%

%
\end{document}